\documentclass[12pt,preprint]{aastex63}

\usepackage{amssymb}
\usepackage{amsmath}
\usepackage[T1]{tipa}
\usepackage{color}
\usepackage{graphicx}
\usepackage{hyperref}
\usepackage{natbib}
\usepackage[capitalise]{cleveref}
\usepackage{multirow}
\usepackage{threeparttablex}
\usepackage{booktabs}
\usepackage{marginnote}
\usepackage{blkarray}
\usepackage{verbatim}
\usepackage{makecell}
\usepackage{wrapfig}

\DeclareTextFontCommand{\textipa}{%
  \fontfamily{cmss}\tipaencoding
}

\newcommand{\kkoname}{k'ni\textipa{P}atn k'l$\left._\mathrm{\smile}\right.$stk'masqt}

\newcommand{\mat}[1]{\mathrm{#1}}
\newcommand{\SNR}{\ensuremath{\text{S/N}}}
\newcommand{\vecb}[1]{\ensuremath{\mathbf{#1}}}

\begin{document}

\title{CHIME/FRB Outriggers: KKO Station System and Commissioning Results}

\shorttitle{KKO Overview}
\shortauthors{CHIME/FRB Collaboration: A.~Lanman, \emph{et al.}}

\author[0000-0003-2116-3573]{Adam E.~Lanman} \altaffiliation{Following the lead, authors are binned into three alphabetized lists of: substantial contributors to analysis of KKO commissioning data, substantial contributors to creation of a functioning instrument at KKO, and other contributors to the success of this project.}
  \affiliation{Department of Physics, McGill University, 3600 rue University, Montr\'eal, QC H3A 2T8, Canada}
  \affiliation{Trottier Space Institute, McGill University, 3550 rue University, Montr\'eal, QC H3A 2A7, Canada}
  \affiliation{MIT Kavli Institute for Astrophysics and Space Research, Massachusetts Institute of Technology, 77 Massachusetts Ave, Cambridge, MA 02139, USA}
  \affiliation{Department of Physics, Massachusetts Institute of Technology, 77 Massachusetts Ave, Cambridge, MA 02139, USA}
\author[0000-0002-3980-815X]{Shion Andrew}
  \affiliation{MIT Kavli Institute for Astrophysics and Space Research, Massachusetts Institute of Technology, 77 Massachusetts Ave, Cambridge, MA 02139, USA}
\author[0000-0002-5857-4264]{Mattias Lazda}
  \affiliation{David A.~Dunlap Department of Astronomy \& Astrophysics, University of Toronto, 50 St.~George Street, Toronto, ON M5S 3H4, Canada}
  \affiliation{Dunlap Institute for Astronomy \& Astrophysics, University of Toronto, 50 St.~George Street, Toronto, ON M5S 3H4, Canada}
\author[0000-0002-4823-1946]{Vishwangi Shah}
  \affiliation{Department of Physics, McGill University, 3600 rue University, Montr\'eal, QC H3A 2T8, Canada}
  \affiliation{Trottier Space Institute, McGill University, 3550 rue University, Montr\'eal, QC H3A 2A7, Canada}
\author[0000-0001-6523-9029]{Mandana Amiri}
  \affiliation{Department of Physics and Astronomy, University of British Columbia, 6224 Agricultural Road, Vancouver, BC V6T 1Z1 Canada}
\author[0000-0003-0477-7645]{Arvind Balasubramanian}
  \affiliation{Department of Astronomy and Astrophysics, Tata Institute of Fundamental Research, Mumbai, 400005, India}
\author[0000-0003-3772-2798]{Kevin Bandura}
  \affiliation{Lane Department of Computer Science and Electrical Engineering, 1220 Evansdale Drive, PO Box 6109, Morgantown, WV 26506, USA}
  \affiliation{Center for Gravitational Waves and Cosmology, West Virginia University, Chestnut Ridge Research Building, Morgantown, WV 26505, USA}
\author[0000-0001-8537-9299]{P.~J.~Boyle}
  \affiliation{Department of Physics, McGill University, 3600 rue University, Montr\'eal, QC H3A 2T8, Canada}
  \affiliation{Trottier Space Institute, McGill University, 3550 rue University, Montr\'eal, QC H3A 2A7, Canada}
\author[0000-0002-1800-8233]{Charanjot Brar}
  \affiliation{Department of Physics, McGill University, 3600 rue University, Montr\'eal, QC H3A 2T8, Canada}
  \affiliation{Trottier Space Institute, McGill University, 3550 rue University, Montr\'eal, QC H3A 2A7, Canada}
\author[0009-0001-7664-5142]{Mark Carlson}
  \affiliation{Department of Physics and Astronomy, University of British Columbia, 6224 Agricultural Road, Vancouver, BC V6T 1Z1 Canada}
\author[0000-0001-6509-8430]{Jean-François Cliche}
  \affiliation{Department of Physics, McGill University, 3600 rue University, Montr\'eal, QC H3A 2T8, Canada}
  \affiliation{Trottier Space Institute, McGill University, 3550 rue University, Montr\'eal, QC H3A 2A7, Canada}
\author[0000-0001-6128-3735]{Nina Gusinskaia}
  \affiliation{Dunlap Institute for Astronomy \& Astrophysics, University of Toronto, 50 St.~George Street, Toronto, ON M5S 3H4, Canada}
  \affiliation{David A.~Dunlap Department of Astronomy \& Astrophysics, University of Toronto, 50 St.~George Street, Toronto, ON M5S 3H4, Canada}
  \affiliation{Canadian Institute for Theoretical Astrophysics, 60 St.~George Street, Toronto, ON M5S 3H8, Canada}
\author[0009-0003-3736-2080]{Ian T.~Hendricksen}
  \affiliation{Department of Physics, McGill University, 3600 rue University, Montr\'eal, QC H3A 2T8, Canada}
  \affiliation{Trottier Space Institute, McGill University, 3550 rue University, Montr\'eal, QC H3A 2A7, Canada}
\author[0000-0003-4810-7803]{J.~F.~Kaczmarek}
  \affiliation{CSIRO Space \& Astronomy, Parkes Observatory, P.O. Box 276, Parkes NSW 2870, Australia}
  \affiliation{Dominion Radio Astrophysical Observatory, Herzberg Research Centre for Astronomy and Astrophysics, National Research Council Canada, PO Box 248, Penticton, BC V2A 6J9, Canada}
\author[0000-0003-1455-2546]{Tom Landecker}
  \affiliation{Dominion Radio Astrophysical Observatory, Herzberg Research Centre for Astronomy and Astrophysics, National Research Council Canada, PO Box 248, Penticton, BC V2A 6J9, Canada}
\author[0000-0002-4209-7408]{Calvin Leung}
  \affiliation{Department of Astronomy, University of California Berkeley, Berkeley, CA 94720, USA}
  \affiliation{NASA Hubble Fellowship Program~(NHFP) Einstein Fellow}
\author[0000-0001-7348-6900]{Ryan Mckinven}
  \affiliation{Department of Physics, McGill University, 3600 rue University, Montr\'eal, QC H3A 2T8, Canada}
  \affiliation{Trottier Space Institute, McGill University, 3550 rue University, Montr\'eal, QC H3A 2A7, Canada}
\author[0000-0002-0772-9326]{Juan Mena-Parra}
  \affiliation{Dunlap Institute for Astronomy \& Astrophysics, University of Toronto, 50 St.~George Street, Toronto, ON M5S 3H4, Canada}
  \affiliation{David A.~Dunlap Department of Astronomy \& Astrophysics, University of Toronto, 50 St.~George Street, Toronto, ON M5S 3H4, Canada}
\author[0000-0001-8292-0051]{Nikola Milutinovic}
  \affiliation{Department of Physics and Astronomy, University of British Columbia, 6224 Agricultural Road, Vancouver, BC V6T 1Z1 Canada}
\author[0000-0003-0510-0740]{Kenzie Nimmo}
  \affiliation{MIT Kavli Institute for Astrophysics and Space Research, Massachusetts Institute of Technology, 77 Massachusetts Ave, Cambridge, MA 02139, USA}
\author[0000-0002-8912-0732]{Aaron B.~Pearlman}
  \affiliation{Department of Physics, McGill University, 3600 rue University, Montr\'eal, QC H3A 2T8, Canada}
  \affiliation{Trottier Space Institute, McGill University, 3550 rue University, Montr\'eal, QC H3A 2A7, Canada}
  \affiliation{Banting Fellow}
  \affiliation{McGill Space Institute Fellow}
  \affiliation{FRQNT Postdoctoral Fellow}
\author[0000-0003-3463-7918]{Andre Renard}
  \affiliation{Dunlap Institute for Astronomy \& Astrophysics, University of Toronto, 50 St.~George Street, Toronto, ON M5S 3H4, Canada}
\author[0000-0003-1842-6096]{Mubdi Rahman}
  \affiliation{Sidrat Research, 124 Merton Street, Suite 507, Toronto, ON M4S 2Z2, Canada}
\author[0000-0002-4543-4588]{J.~Richard Shaw}
  \affiliation{Department of Physics and Astronomy, University of British Columbia, 6224 Agricultural Road, Vancouver, BC V6T 1Z1 Canada}
\author[0000-0003-2631-6217]{Seth R.~Siegel}
  \affiliation{Perimeter Institute for Theoretical Physics, 31 Caroline Street N, Waterloo, ON N25 2YL, Canada}
  \affiliation{Department of Physics, McGill University, 3600 rue University, Montr\'eal, QC H3A 2T8, Canada}
  \affiliation{Trottier Space Institute, McGill University, 3550 rue University, Montr\'eal, QC H3A 2A7, Canada}
\author[0000-0002-6873-2094]{Rick J.~Smegal}
  \affiliation{Department of Physics and Astronomy, University of British Columbia, 6224 Agricultural Road, Vancouver, BC V6T 1Z1 Canada}
\author[0000-0003-2047-5276]{Tomas Cassanelli}
  \affiliation{Department of Electrical Engineering, Universidad de Chile, Av. Tupper 2007, Santiago 8370451, Chile}
\author[0000-0002-2878-1502]{Shami Chatterjee}
  \affiliation{Cornell Center for Astrophysics and Planetary Science, Cornell University, Ithaca, NY 14853, USA}
\author[0000-0002-8376-1563]{Alice P.~Curtin}
  \affiliation{Department of Physics, McGill University, 3600 rue University, Montr\'eal, QC H3A 2T8, Canada}
  \affiliation{Trottier Space Institute, McGill University, 3550 rue University, Montr\'eal, QC H3A 2A7, Canada}
\author[0000-0001-7166-6422]{Matt Dobbs}
  \affiliation{Department of Physics, McGill University, 3600 rue University, Montr\'eal, QC H3A 2T8, Canada}
  \affiliation{Trottier Space Institute, McGill University, 3550 rue University, Montr\'eal, QC H3A 2A7, Canada}
\author[0000-0003-4098-5222]{Fengqiu Adam Dong}
  \affiliation{Department of Physics and Astronomy, University of British Columbia, 6224 Agricultural Road, Vancouver, BC V6T 1Z1 Canada}
\author[0000-0002-1760-0868]{Mark Halpern}
  \affiliation{Department of Physics and Astronomy, University of British Columbia, 6224 Agricultural Road, Vancouver, BC V6T 1Z1 Canada}
\author[0009-0002-1199-8876]{Hans Hopkins}
  \affiliation{Perimeter Institute for Theoretical Physics, 31 Caroline Street N, Waterloo, ON N25 2YL, Canada}
\author[0000-0001-9345-0307]{Victoria M.~Kaspi}
  \affiliation{Department of Physics, McGill University, 3600 rue University, Montr\'eal, QC H3A 2T8, Canada}
  \affiliation{Trottier Space Institute, McGill University, 3550 rue University, Montr\'eal, QC H3A 2A7, Canada}
\author[0009-0005-7115-3447]{Kholoud Khairy}
  \affiliation{Lane Department of Computer Science and Electrical Engineering, 1220 Evansdale Drive, PO Box 6109, Morgantown, WV 26506, USA}
  \affiliation{Center for Gravitational Waves and Cosmology, West Virginia University, Chestnut Ridge Research Building, Morgantown, WV 26505, USA}
\author[0000-0002-4279-6946]{Kiyoshi W.~Masui}
  \affiliation{MIT Kavli Institute for Astrophysics and Space Research, Massachusetts Institute of Technology, 77 Massachusetts Ave, Cambridge, MA 02139, USA}
  \affiliation{Department of Physics, Massachusetts Institute of Technology, 77 Massachusetts Ave, Cambridge, MA 02139, USA}
\author[0000-0001-8845-1225]{Bradley W.~Meyers}
  \affiliation{International Centre for Radio Astronomy Research, Curtin University, Bentley, WA 6102, Australia}
\author[0000-0002-2551-7554]{Daniele Michilli}
  \affiliation{MIT Kavli Institute for Astrophysics and Space Research, Massachusetts Institute of Technology, 77 Massachusetts Ave, Cambridge, MA 02139, USA}
  \affiliation{Department of Physics, Massachusetts Institute of Technology, 77 Massachusetts Ave, Cambridge, MA 02139, USA}
\author[0000-0002-9822-8008]{Emily Petroff}
  \affiliation{Perimeter Institute for Theoretical Physics, 31 Caroline Street N, Waterloo, ON N25 2YL, Canada}
\author[0000-0002-9516-3245]{Tristan Pinsonneault-Marotte}
  \affiliation{Department of Physics and Astronomy, University of British Columbia, 6224 Agricultural Road, Vancouver, BC V6T 1Z1 Canada}
\author[0000-0002-4795-697X]{Ziggy Pleunis}
  \affiliation{Dunlap Institute for Astronomy \& Astrophysics, University of Toronto, 50 St.~George Street, Toronto, ON M5S 3H4, Canada}
\author[0000-0001-7694-6650]{Masoud Rafiei-Ravandi}
  \affiliation{Department of Physics, McGill University, 3600 rue University, Montr\'eal, QC H3A 2T8, Canada}
  \affiliation{Trottier Space Institute, McGill University, 3550 rue University, Montr\'eal, QC H3A 2A7, Canada}
\author[0000-0002-6823-2073]{Kaitlyn Shin}
  \affiliation{MIT Kavli Institute for Astrophysics and Space Research, Massachusetts Institute of Technology, 77 Massachusetts Ave, Cambridge, MA 02139, USA}
  \affiliation{Department of Physics, Massachusetts Institute of Technology, 77 Massachusetts Ave, Cambridge, MA 02139, USA}
\author[0000-0002-2088-3125]{Kendrick Smith}
  \affiliation{Perimeter Institute for Theoretical Physics, 31 Caroline Street N, Waterloo, ON N25 2YL, Canada}
\author[0000-0003-4535-9378]{Keith Vanderlinde}
  \affiliation{David A.~Dunlap Department of Astronomy \& Astrophysics, University of Toronto, 50 St.~George Street, Toronto, ON M5S 3H4, Canada}
  \affiliation{Dunlap Institute for Astronomy \& Astrophysics, University of Toronto, 50 St.~George Street, Toronto, ON M5S 3H4, Canada}
\author[0000-0002-7076-8643]{Tarik J.~Zegmott}
  \affiliation{Department of Physics, McGill University, 3600 rue University, Montr\'eal, QC H3A 2T8, Canada}
  \affiliation{Trottier Space Institute, McGill University, 3550 rue University, Montr\'eal, QC H3A 2A7, Canada}
\newcommand{\allacks}{
A.B.P. is a Banting Fellow, a McGill Space Institute~(MSI) Fellow, and a Fonds de Recherche du Quebec -- Nature et Technologies~(FRQNT) postdoctoral fellow.
A.P.C is a Vanier Canada Graduate Scholar
C. L. is supported by NASA through the NASA Hubble Fellowship grant HST-HF2-51536.001-A awarded by the Space Telescope Science Institute, which is operated by the Association of Universities for Research in Astronomy, Inc., under NASA contract NAS5-26555.
F.A.D is supported by the UBC Four Year Fellowship
K.M.B. is supported by NSF grant 2018490
K.N. is an MIT Kavli fellow. 
K.S.\ is supported by the NSF Graduate Research Fellowship Program.
K.W.M. holds the Adam J. Burgasser Chair in Astrophysics and is supported by NSF grants (2008031, 2018490).
M.D. is supported by a CRC Chair, NSERC Discovery Grant, CIFAR, and by the FRQNT Centre de Recherche en Astrophysique du Qu\'ebec (CRAQ).
V.M.K. holds the Lorne Trottier Chair in Astrophysics \& Cosmology, a Distinguished James McGill Professorship, and receives support from an NSERC Discovery grant (RGPIN 228738-13), from an R. Howard Webster Foundation Fellowship from CIFAR, and from the FRQNT CRAQ.
Z. P. is a Dunlap Fellow.
}

\correspondingauthor{Adam E. Lanman}
\email{alanman@mit.edu}

\begin{abstract}
	Localizing fast radio bursts (FRBs) to their host galaxies is an essential step to better understanding their origins and using them as cosmic probes. The CHIME/FRB Outrigger program aims to add VLBI-localization capabilities to CHIME, such that FRBs may be localized to tens of milliarcsecond precision at the time of their discovery, more than sufficient for host galaxy identification. The first-built outrigger telescope is the \kkoname{} outrigger (KKO), located 66 kilometers west of CHIME. Cross-correlating KKO with CHIME can achieve arcsecond precision along the baseline axis while avoiding the worst effects of the ionosphere. Since the CHIME --- KKO baseline is mostly East/West, this improvement is mostly in right ascension. This paper presents measurements of KKO's performance throughout its commissioning phase, as well as a summary of its design and function. We demonstrate KKO's capabilities as a standalone instrument by producing full-sky images, mapping the angular and frequency structure of the primary beam, and measuring feed positions. To demonstrate the localization capabilities of the CHIME --- KKO baseline, we collected five separate observations each for a set of twenty bright pulsars, and aimed to measure their positions to within 5~arcseconds. All of these pulses were successfully localized to within this specification. The next two outriggers are expected to be commissioned in 2024, and will enable subarcsecond localizations for approximately hundreds of FRBs each year.
\end{abstract}

\section{Introduction}
Fast radio bursts (FRBs) are micro/millisecond pulses of radio emission observed with extreme dispersion, implying an extragalactic origin \citep{lbm+07}. The association of 30 FRBs (to date) with host galaxies has shown this conclusively, implying highly energetic origins. The exact nature of the FRB sources remains elusive, since FRB properties alone have thus far been insufficient to distinguish among the many plausible models for their origins (see \cite{phl22} for a review). Understanding the nature of FRBs, in addition to solving an intriguing cosmic mystery, will also better enable them to be used as probes of the circumgalactic and intergalactic media (CGM \& IGM, respectively), and constrain models of galaxy evolution and cosmology.

Identifying and characterizing the host galaxies of FRB sources and studying their local environments are thus the next major steps in FRB science. At arcsecond resolution, hosts with nearly Milky Way luminosity can be identified for FRBs out to about $z \sim 0.1$ \citep{eb17}. The limited sample of identified host galaxies, such as those from the Australian SKA Pathfinder (ASKAP) and Very Large Array (VLA), has already greatly constrained FRB progenitors. Surveys of FRB host galaxies have generally found a tendency for FRBs to occur in star-forming galaxies, but a sizeable fraction are also found in more quiescent galaxies \citep{Heintz2020, bha+22, gordon_demographics_2023}. At higher angular resolution, offsets between FRB locations and galactic centers can be measured. \cite{mfs+21} compared the distribution of offsets to other transients, disfavoring long gamma-ray burst models and favoring formation channels similar to core-collapse supernovae and binary neutron star mergers. The Deep Synoptic Array 110 (DSA-110) recently put out its first uniform sample of FRB host galaxies, which the authors argue supports progenitor formation channels associated with old stellar populations \citep{law_deep_2023, ravi_deep_2023}. Sub-kpc scale localization has been achieved with very long baseline interferometry (VLBI) observations of some \emph{repeating} FRBs, showing a variety of host environments. European VLBI Network (EVN) observations of the periodic FRB 20180916B showed it to be offset from a nearby region of star formation in its host \citep{mnh+20,tendulkar_60_2021}, while observations of FRB 20201124A found it to be in a star-forming region of its host galaxy \citep{nimmo_milliarcsecond_2021}. In contrast, EVN observations of FRB 20200120E found it to be in a globular cluster of M81 \citep{kmn+22}, an environment with very little star formation.

A great deal of diversity in local environments has thus been identified, even with a limited sample of FRB host galaxies, which may indicate a variety of source classes. A much larger sample of localized FRBs is required to better explore this parameter space, but obtaining such a sample is instrumentally challenging. With higher resolution, more beams must be formed to cover a given field of view, increasing the computational complexity of the search. Searching a narrower field of view solves this problem, but reduces the overall rate of FRB detection.

The Canadian Hydrogen Intensity Mapping Experiment (CHIME) FRB backend is currently the world's leading FRB detector, discovering an average of three new FRBs every day \citep{2018ApJ...863...48C}. By writing out buffered baseband data  on a low-latency trigger, CHIME/FRB can reach its theoretical limit in angular resolution of around an arcminute \citep{mbb+22} This has enabled host galaxy identification for FRBs within 300~Mpc \citep{bhardwaj_host_2023, ibik2023proposed, michilli2023subarcminute, bgk+21, bkm+21} but is insufficient for localizing more distant FRBs.

To overcome this limitation, the CHIME/FRB Outriggers program is deploying three new CHIME-like outrigger telescopes across North America to enable very-long baseline interferometry (VLBI). The full Outriggers program will be discussed in a forthcoming project overview paper. The first of these outriggers, \kkoname{} (KKO), saw first light in June of 2022 and reached full science operations in September of 2023. The second outrigger is located in the Green Bank Observatory (GBO) in West Virginia, began commissioning in April 2023, and the third outrigger site, in the Hat Creek Observatory in northern California, is expected to begin operations in June of 2024. Each outrigger is a single CHIME cylinder, rotated and rolled such that its field of view overlaps with the CHIME field of view. Each outrigger maintains a 40~s buffer of baseband data, which it will write to disk upon receiving an FRB detection trigger from CHIME. The data are then beamformed and cross-correlated offline. This triggered-VLBI method for transient localization has been demonstrated using pathfinder outrigger telescopes \citep{clr+21, lmm+21,slb+23,cls+23}.

KKO is located about 66~km from CHIME, forming the shortest baseline in the outrigger network. At this length, the CHIME --- KKO baseline can achieve arcsecond-scale resolution while avoiding the worst effects of the ionosphere. This paper gives a complete description of KKO design (Section~\ref{sec:design}), and measurements of its performance in commissioning as both a standalone telescope (Section~\ref{s:standalone}) and a VLBI station (Section~\ref{sec:vlbi}). We demonstrate the VLBI capabilities of the KKO --- CHIME baseline by localizing, to 4~arcsec precision, several single pulses from 20 pulsars whose astrometric positions are known at the sub-arcsecond level.

\section{Design Overview}
\label{sec:design}

The KKO system is similar to CHIME in terms of its optical design, hardware, and software subsystems. In this paper we will focus on differences with CHIME, referring the interested reader to \citet{2022ApJS..261...29C} (hereafter, the CHIME Overview). The biggest difference is the scale of the system, which has 1/16 of CHIME's collecting area. The later stages of the data processing also differ, reflecting KKO's sole science goal of transient VLBI. Vital
characteristics of the system can be found in Table~\ref{t:characteristics}.

\begin{table}
    \centering
    \begin{tabular}{l l}
        \hline
        Geographic location &  49$^\circ$ 25' 8.5794'' N\\
                            &  120$^\circ$ 31' 31.08'' W \\
        Elevation & 800\,m \\
        Baseline to CHIME$^a$ & 66.2\,km \\
        Cylinder width & 20\,m \\
        Cylinder length & 40\,m \\
        Instrumented cylinder length & 19.5\,m \\
        Number of dual-polarization feeds & 64 \\
        Observing band & 400--800\,MHz \\
        \hline
    \end{tabular}
    \caption{\label{t:characteristics}
    Characteristics of KKO.
    Notes: (a) Projected baseline for the target at the CHIME zenith.
    }
\end{table}

\subsection{Site}
\label{s:site}
\begin{figure}
	\centering
	\includegraphics[width=\linewidth]{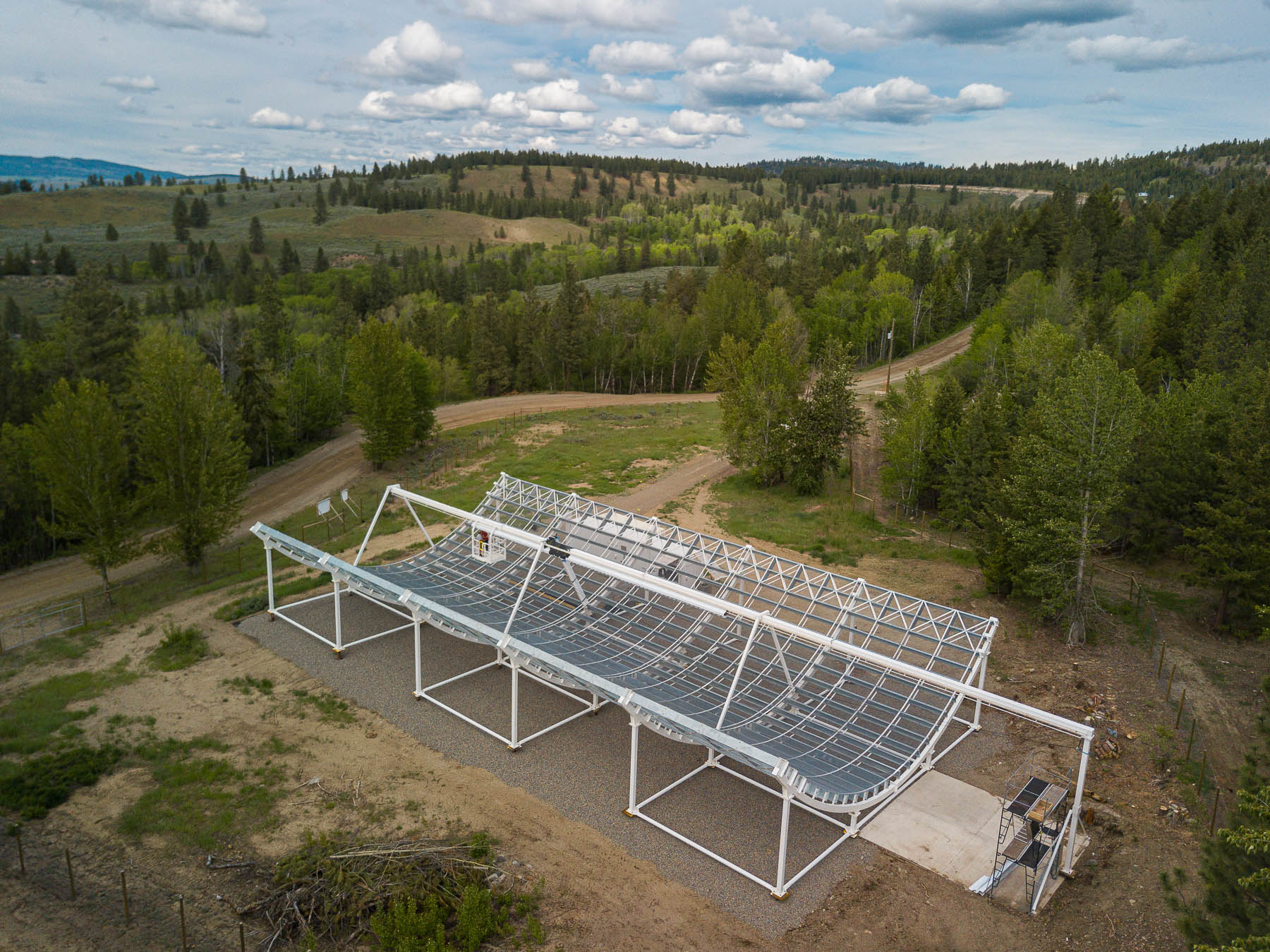}
	\includegraphics[width=\linewidth]{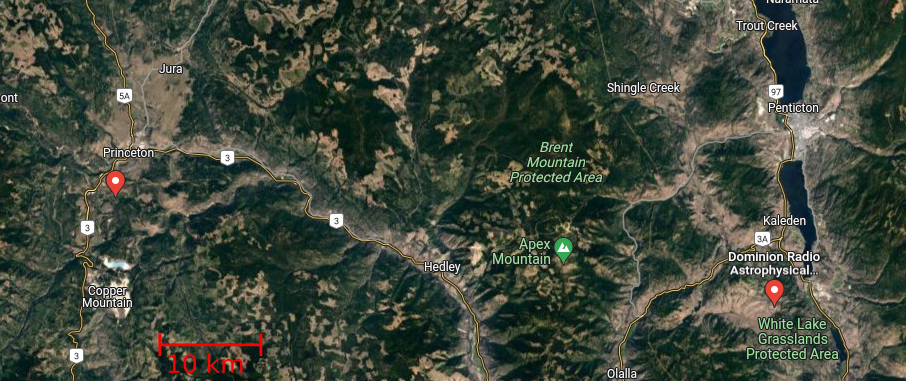}
    \caption{The KKO site. \textsl{Top:} Photo of KKO and its site. This photo was taken during servicing, with the service trolley occupied and shifted over the cylinder. During normal operations, the trolley is kept by the scaffolding to the right. \textsl{Bottom:}  Map showing the KKO geographic location relative to CHIME and nearby population centers. North is up. Satellite imagery \copyright 2023 TerraMetrics (2023), Map data \copyright 2023 Google}
    \label{fig:site}
\end{figure}

The KKO site is located 4~kilometers South of Princeton, BC, on the North slopes of Agate and Copper Mountains. With a $\sim 66$~km line-of-sight distance to CHIME, it is far enough to provide a useful VLBI baseline while being close enough for CHIME staff at UBC and Dominion Radio Astrophysical Observatory (DRAO) to perform maintenance as needed. Figure~\ref{fig:site} shows a photo of the KKO site and a map of its location relative to the town of Princeton and to CHIME.

The site is on the traditional unceded territory of the Similamix People. We communicated with the Upper Similkameen Band before and during the construction. They have generously offered the name for the telescope, which in Upper Similkameen means ``a listening device for outer space.'' The parcel of land occupied by the outrigger is leased from Heritage Mining INC. Co-located with CHIME on the Southern end of the Thompson Plateau of British Columbia, the KKO site shares similar geomorphology and semi-arid climate. However, being $\sim 200$~m higher (at 800~m above sea level), it receives more precipitation. For comparison, the annual snowfall at Princeton is often double that of Penticton.

The site offers excellent RFI protection from the South and the neighbouring households to the East, but it is exposed in the direct line of sight to the town towards the North. A significant portion of the KKO's frequency band is contaminated by wireless communication, TV broadcasting bands, direct transmission from satellites and airplanes, as well as the scattering of distant ground-based sources. Unfortunately, a new LTE/5G tower was installed on the nearby mountain to the northwest during the instrument's construction. This caused a significant increase in RFI power level, in the band from 700 to 760 MHz, which prompted a redesign of the analog chain, culminating in the insertion of analog band-stop filters in LTE frequencies.

\subsection{Cylindrical Reflector}
\label{s:structure}

The telescope reflector is functionally identical to CHIME; It is a cylindrical paraboloid 20 meters wide and 40 meters long, with a 5 meter focal length. The central 20 meters of the focal line is fitted with a linear focal-plane array of feeds and receivers. Its long axis is aligned approximately North-South, rotated from the NS line and tipped slightly to the East so that the telescope sees exactly the same sky as CHIME. Because the line-of-sight separation is only 66~km, the rotation of 0.69$^\circ$ and tilt of 0.59$^\circ$, are not apparent to the naked eye.  Three support posts for the focal line are placed where the structural design requires them, two of which are within the working aperture. Their locations do not match those of equivalent structures on CHIME, but the effect on the telescope response is minimal.

The site is not naturally level; it was levelled about one hundred years ago, and used as a railway siding. The mine tailings used to level the site consist of very small particles, which yield very poor bearing capacity. To provide foundations for the telescope, 24 helical pilings, 30~cm diameter steel tubes, were driven into the ground to depths of 2 to 7 meters.  A grillage of steel pipes connects the tops of the pilings into a single structure that supports the telescope.

The reflector was designed by Sightline Engineering in Vernon, B.C., for a snow load heavier than that at CHIME, so that steel structural members are of higher gauge material and more prominent in size. In addition, extra strapping was installed under the mesh to support the heavier snow load and compensate for the smaller size of the reflector mesh panels. The steel structure matches the design dimensions to a tolerance of 5~mm. The reflector mesh is galvanized steel with 19~mm (3/4 inch) openings. The mesh is shaped to a surface precision of order 1~cm.

The most significant departure from the CHIME design is that access to the KKO focus is from below, not above. The top of the focal line is closed, effectively keeping water out of focal-line equipment. Access to the focus is slightly less convenient, gained from a travelling cart that runs on two rails fixed to the sides of the focal-line structure. The cart, made of welded steel, can carry two people, who move it by pulling on a rope.

\subsection{Receiver Chain}
\label{s:analog}

\begin{figure}[t]
	\includegraphics[width=\linewidth]{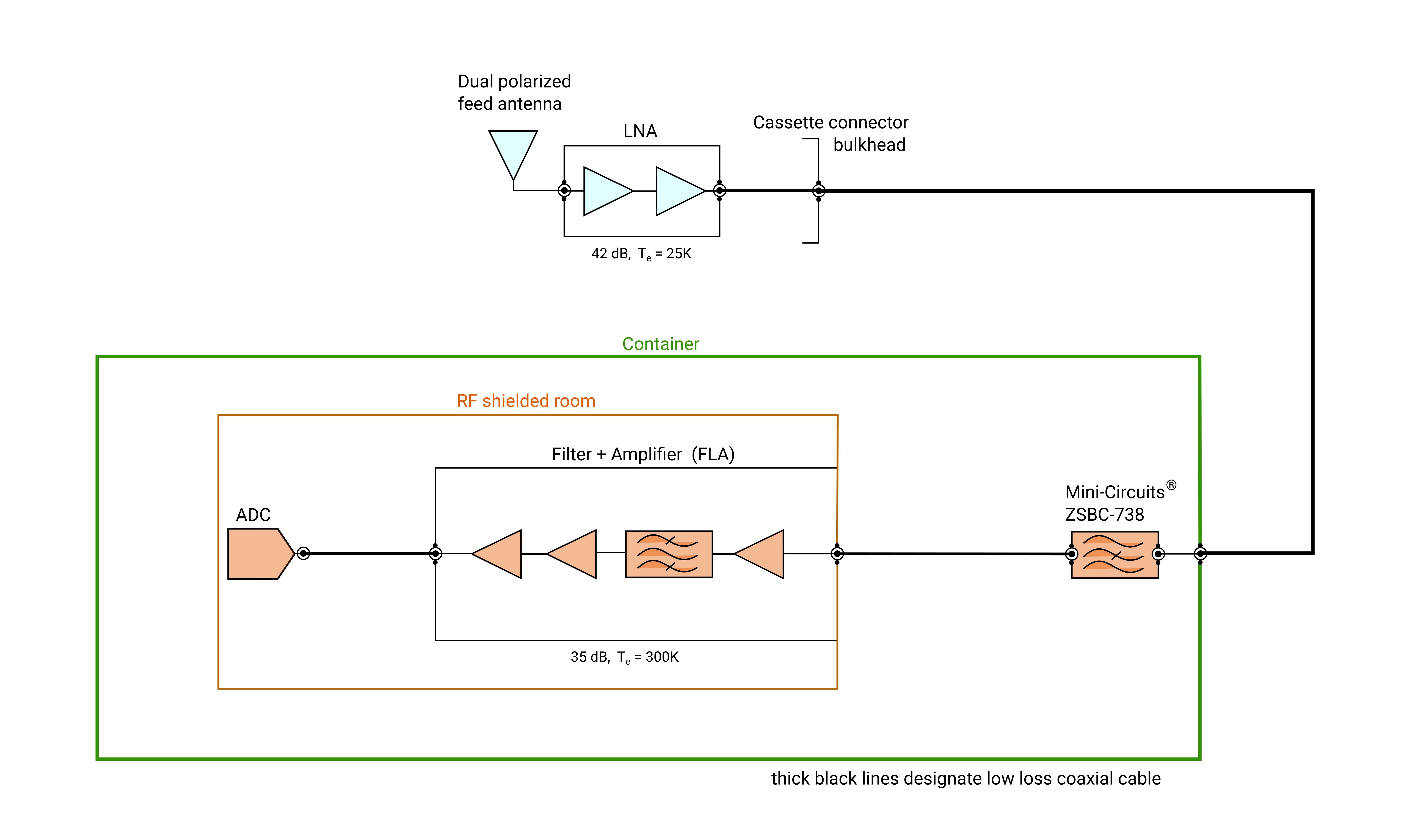}
	\caption{\label{f:Receiver Chain} The analog system from the antenna to the ADC. The antenna feed and LNA are contained in an enclosed cassette on the focal line, which protects them from weather. Low-loss coaxial cable connects the casette to the the exterior of the receiver hut, which feeds through to an inner bulkhead which holds the Mini-Circuits custom LTE filter. The LTE-filtered signal is then passed into the RF-shielded room, amplified by the FLA, and sampled.}
	\label{fig:analog_chain}
\end{figure}

Figure~\ref{fig:analog_chain} provides a block diagram of the analog receiver chain. The design of the analog receiver chain -- dual-polarization antenna feed, amplifiers, gain distribution,  and filtering -- essentially duplicates that of CHIME and is described in detail in the CHIME Overview. There is no frequency conversion: signals in the 400 to 800 MHz band are amplified by low-noise amplifiers (LNAs) at the feeds and further amplified in second-stage filter-amplifiers (FLAs) installed in a shielded room in the receiver hut. The ability to replicate an existing proven design with established manufacturing and test processes was a key to the fast deployment of the receiver chain in the field and, at the same time, significant cost saving on the engineering front. Sourcing components during supply-chain issues caused by the global pandemic at the time (2020 -- 2021) was challenging. This was especially true for sourcing the discontinued low-noise transistor from Avago, which is critical to CHIME's excellent noise performance.

Feed spacing along the focal line is 305~mm, identical to CHIME. Two feeds, each receiving two orthogonal linear polarizations, are mounted in a cassette. Each feed is protected by a radome, vacuum-formed from 0.090" ABS polymer into a close-fitting cover. Cassettes are assembled from aluminum components made by water-jet cutting and bending. The two linear polarizations are aligned along the focal line (Y) and perpendicular to the focal line (X).

Each feed is equipped with two LNAs, one per polarization. This LNA has a gain of 42~dB with a roll-off above 1~GHz. As in CHIME, DC power for LNAs is supplied through the coaxial cables, which carry the amplified signals off the focal line. This way, electronic equipment at the focal line is kept at the absolute minimum. Noise performance across the 400 -- 800~MHz band at KKO is comparable to that at CHIME. Cables that carry signals from the focal line to the shielded room are 35~m long, LMR-400 type, and manufactured at the same electrical length within 25~cm. The cables are protected from the weather throughout their path to the receiver hut by covers along the focal line, in a closed conduit down the focal-line support, and covered cable trays underneath the reflector. This protection is essential in maintaining phase equality on all signal paths.

Signals enter the receiver hut through a bulkhead and pass through bandstop filters, designed to block LTE signals (see Section~\ref{sec:lte_filt} for details of the design of these filters). After entering a shielded room with a shielding effectiveness of 100 dB, signals pass to second-stage Filter amplifiers (FLA)s, with 35~dB of gain. FLAs include a bandpass filter (400 to 800 MHz) to define the CHIME band and enable direct sampling in the second Nyquist zone. FLAs also feed DC to the LNAs on the focal line. The DC supply current for each FLA/LNA pair passes through a TPS259620 electronic fuse from Texas Instruments to provide overload protection. Finally, output levels from FLAs are set by fixed attenuators to values appropriate for the analog to digital converters (ADCs).

Each signal passes through eleven coaxial connectors between the feed and the ADC. The use of connectors adds cost and may create reliability problems, but it greatly simplifies testing, modular assembly and maintenance of the system.

\subsubsection{Focal-line feed assembly}

Cassettes are supported by a frame of square steel tubes welded to the focal-line structure. Assembled cassettes, each holding two feeds and their LNAs, and fitted with their radomes, weigh 3.5~kg.  Brackets to join them to the support frame are formed from aluminum sheet metal, cut to shape with a water jet. The ground plane for the focal line is made continuous by joining each cassette to its neighbour with an aluminum strip. Coaxial connectors (Type N) on the outside of each cassette connect to 35-m cables, which lie above the cassettes, supported by thin sheets of corrugated plastic roofing material. Two formed roofs of painted sheet steel are fixed on either side of the main structure of the focal line as a complete shield against rain and snow melt for the cassettes. Coaxial cables pass down a ``chimney'', a square sheet metal conduit strapped to one of the legs that supports the focal line. These cables are exposed on top of the focal line for about 1.5m and protected from snow and rain by a sheet-metal roof.

\subsubsection{LTE bandstop filters and attenuation}
\label{sec:lte_filt}

Between the selection of the site in 2019 and the first operation of the telescope in 2022, a new communications transmitter was installed in the area, a few km North of KKO. Signal levels in the LTE band, 710 to 757 MHz, became unacceptably high. In response, we installed a custom bandstop filter ZSBC-738, designed to our specifications by Mini-Circuits\footnote{\url{www.minicircuits.com}} to manage signal level at the input of the ADC. The design is a compromise between eliminating the LTE signal altogether and retaining as much useful bandwidth as possible (especially important near the top of the band). The signal was attenuated to a level where the analog and digital signal processing could cope without degrading the telescope's sensitivity. The final design achieves attenuation greater than 15~dB between 718 and 758~MHz, and allows us to make good use of all frequencies outside the range 707 to 762~MHz. The filters are able to carry the LNA supply current of 120 mA, and it was possible to insert them between the LNAs and FLAs.

\subsection{Correlator}
\label{s:correlator}

\subsubsection{F-engine}

The KKO digital backend is designed as an FX correlator. The F-engine is implemented using the same field programmable gate array (FPGA)-based ICE framework used in CHIME \citep{2016JAI.....541005B,2022ApJS..261...29C}. A total of eight ICE motherboards packed in a crate with an ICE backplane perform the sampling, frequency channelization, and corner-turn networking operations of the correlator. The ADC on each ICE board samples the incoming voltage stream at 800~MS/s. Due to the filtering in the FLAs, this recovers the 400 -- 800~MHz operating band of CHIME, corresponding with the second Nyquist zone. Channelization is done with a polyphase filter bank (PFB) in combination with a fast Fourier transform (FFT), producing 1024 channels, each 390~kHz wide. The ICE software and firmware have been upgraded from those at CHIME to enable a corner-turn using half of the boards of a full (16-board) ICE crate.

The design of the F-engine cooling is changed significantly from CHIME. The RF-shielded equipment room is cooled by a conventional 5T York air conditioner (AC). The outdoor component of the AC unit, the condenser, does not use any RF-emitting electronics. Coolant lines pass through the wall of the shielded room via waveguide-pipe penetration, and the control line is filtered using an Astrodyne low-pass filter with 100-dB stopband attenuation.  A portable Koolance EXC-800 chiller located inside the room provides water cooling to radiators on the FPGA crate. Fans drive air over these radiators, providing very stable ADC and FPGA temperatures.

\subsubsection{X-engine}

The second stage of digital hardware is the \emph{X-engine}. While this system is primarily a VLBI baseband recorder, we nonetheless refer to it as the X-engine due to its heritage from the CHIME X-engine and its secondary function of cross-correlating (the X) the data for calibration purposes.

The X-engine is similar in concept to CHIME's X-engine (as described in the CHIME Overview) along with its triggered baseband recording system (see \citet{abb+18}). It receives the corner-turned data from the F-engine, buffers it for triggered baseband recording of VLBI data (see Subsection~\ref{s:baseband}), performs cross-multiplication and integration to produce visibilities, and transmits data products to other machines for storage or further processing. This is done on a platform of x86-architecture computers housing GPUs for accelerated computing. 

The scale of the outrigger X-engine is determined by the throughput of input data, in contrast to CHIME, where it was set by the compute load. This change is a result of two factors -- improvements in GPU processing power (allowing each to handle more frequencies at a time), and the quadratic scaling of compute cost with number of inputs. The KKO X-engine is a 128-input correlator composed of a 2-node cluster, where each node houses four 2$\times$ 40G network cards, 2 GPUs, and 1.5\,TB of memory. Compare this with CHIME's first-light\footnote{We specify first-light because CHIME will soon undergo an update of its X-engine hardware} X-engine, which is a 2048 input correlator composed of a 256-node cluster where each node houses one 4$\times$ 10G network card, 4 GPUs, and 64\,GB of memory. The $N^2$ compute load per node is higher for CHIME, whereas the input data throughput per node is substantially higher for the Outriggers. The full component list for the X-engine nodes is presented in Table~\ref{t:xengine}.

\begin{table}
    \centering
    \begin{tabular}{l l l}
        \hline
        Component & Make and model & Number per node \\
        \hline
        Chassis & Gigabyte G482-Z52 & 1 \\
        Motherboard & Gigabyte MZ52-G40 & 1 \\
        CPU & AMD EPYC 7402 24-Core Processor & 2 \\
        GPU & AMD MI100 & 2 \\
        Input data NIC & 2xQSFP+ ports (operating as 2x 4x10G logical links) & 4\\
        Output data NIC & 2xSFP28 ports (2x25G links) & 1\\
        Memory & Kingston 64~GB and 32~GB & 16 of each\\
        SSDs & Intel D3-S4610 (240 GB) & 1 \\
        \hline
    \end{tabular}
    \caption{\label{t:xengine} Parts list for each of the two X-engine nodes.}
\end{table}

\subsubsection{Computing and Networking}

Data processing after the X-engine is performed on a cluster of three compute/storage nodes. These are 45 Drives Q30 Storinator servers, with dual Intel Xeon 4210 processors, and 125 GB of memory, equipped with ten 6 TB SAS hard drives configured with RAID as a storage system, along with a 500~GB solid-state drive for fast data staging. The computing system also includes two ``auxiliary'' nodes, each a Supermicro brand SuperServer. These manage the distribution of processes across other nodes and run services not on the primary data path, such as collecting housekeeping metrics and handling emergency shutdowns.

Data are transferred over QSFP+ fiber-optic links (operating in a 4x10 Gbps ethernet mode) between the F-engine and the X-engine, and 25 Gbps ethernet connections take data from the X-engine to the receiver (compute) nodes. Standard 1~Gbps ethernet links connect all nodes through two network switches for regular connectivity. A ``gateway'' node is connected to the Internet through the Internet service provided by ChinaCreek ISP via a microwave link to the site.

The core functionality of the ChinaCreek ISP point-to-point Internet air link is provided by an Airspan Mimosa B5c radio operating between 4.9 and 6.2~GHz mounted within a weatherproof and RF-shielded enclosure. Two 5-m long LMR-400 coaxial cables connect the enclosed radio near the base of the pole to a dual-polarized conical horn antenna near the top. This enclosure and the antenna are mounted on a power utility pole. A managed Gb-speed network switch sharing the same enclosure provides data connectivity and 48~V to power the radio via a short power-over-ethernet (PoE) cable. The switch is intended to be powered by batteries and accepts a wide input voltage range (9 to 72~V DC). The switch accepts SFP transceivers, which complete the data connection to the receiver hut over an 80-m long single-mode armored optical fiber contained within a buried PVC conduit. Power to the switch is provided by an uninterruptible power supply (UPS) in the receiver hut and connected via an 80-m long wire pair sharing the conduit with the optical fiber.

A backup connection is enabled through an LTE modem to protect system access during power or Internet outages. The antenna connection to LTE modem, which is located inside of the RF room, is shielded via the high-pass filters that attenuate all the frequencies under 1500~MHz. This way, the modem can access available higher-frequency LTE bands without exposing the feeds to RFI from on-site LTE emissions within the CHIME band. Further, the gateway node, switches, the clock, the enviromux sensor suite, and networking hardware are all connected to a UPS, which can power these devices for up to an hour. This backup connection ensures we can monitor site conditions and take protective actions despite limited access during outages.

\subsection{Clocking system}
\label{s:clocking}

The clock and timing signals for the F-engine are generated by a Spectrum Instruments TM-4 GPS-disciplined crystal oscillator\footnote{\url{https://www.spectruminstruments.net/TM4um.pdf}} and distributed to each motherboard through the ICE backplane. As in CHIME, the stability of this frequency standard is sufficient to drive the data acquisition and correlator operations of KKO as a stand-alone interferometer. However, the GPS oscillator is not sufficiently stable for wide-field VLBI. We implement the clock stabilization system developed by \cite{mlc+22} to transfer the reference clock to a more stable frequency standard during VLBI observations. 

The stabilization system works by injecting an analog signal derived from a more stable clock through one of the correlator inputs. For KKO this stable clock is a free-running Stanford Research Systems FS725 Rubidium clock, while CHIME uses a passive hydrogen maser. In both cases, the F-engine periodically captures a subset of the raw (not frequency-channelized) ADC data from the stable clock input. These data are processed offline to monitor the variations of the GPS clock and correct for the clock-induced phase variations in the baseband data between VLBI calibration observations. By default, the clock stabilization system monitors GPS clock variations at 200~ms cadence, much faster than the characteristic timescale of timing variations in the GPS module ($\gtrsim$1~s), which are dominated by the tuning algorithm that disciplines the internal crystal oscillator of the TM-4. By correcting for the GPS clock variations, the clock timing errors between CHIME and KKO can be maintained below 200~ps on $\sim 10^3$~s timescales, which is enough time to phase reference an FRB observation to a VLBI calibrator.

\subsection{Real-time Data Processing Pipeline}

The two main data pipelines are the \emph{triggered baseband pipeline}, which writes out channelized baseband data directly to disk upon receiving a trigger from CHIME/FRB, and the \emph{$N^2$ visibility pipeline}, which constructs standard interferometric visibilities among the 128 antenna feeds. The baseband pipeline is the source of the main scientific data products, ultimately providing the coherent, high-time-resolution beams to cross-correlate with CHIME in the off-site VLBI correlator. The $N^2$ pipeline exists primarily for gain calibration, and has been useful for characterizing system noise, RFI contamination, and the primary beam shape. Data pipelines after the F-engine are constructed using the \texttt{kotekan} framework \citep{renard_kotekan_2021}, which provides highly-optimized processing stages to perform various computations on interferometric data.

\begin{figure}
	\hspace*{-0.025\linewidth}
	\includegraphics[width=1.05\linewidth]{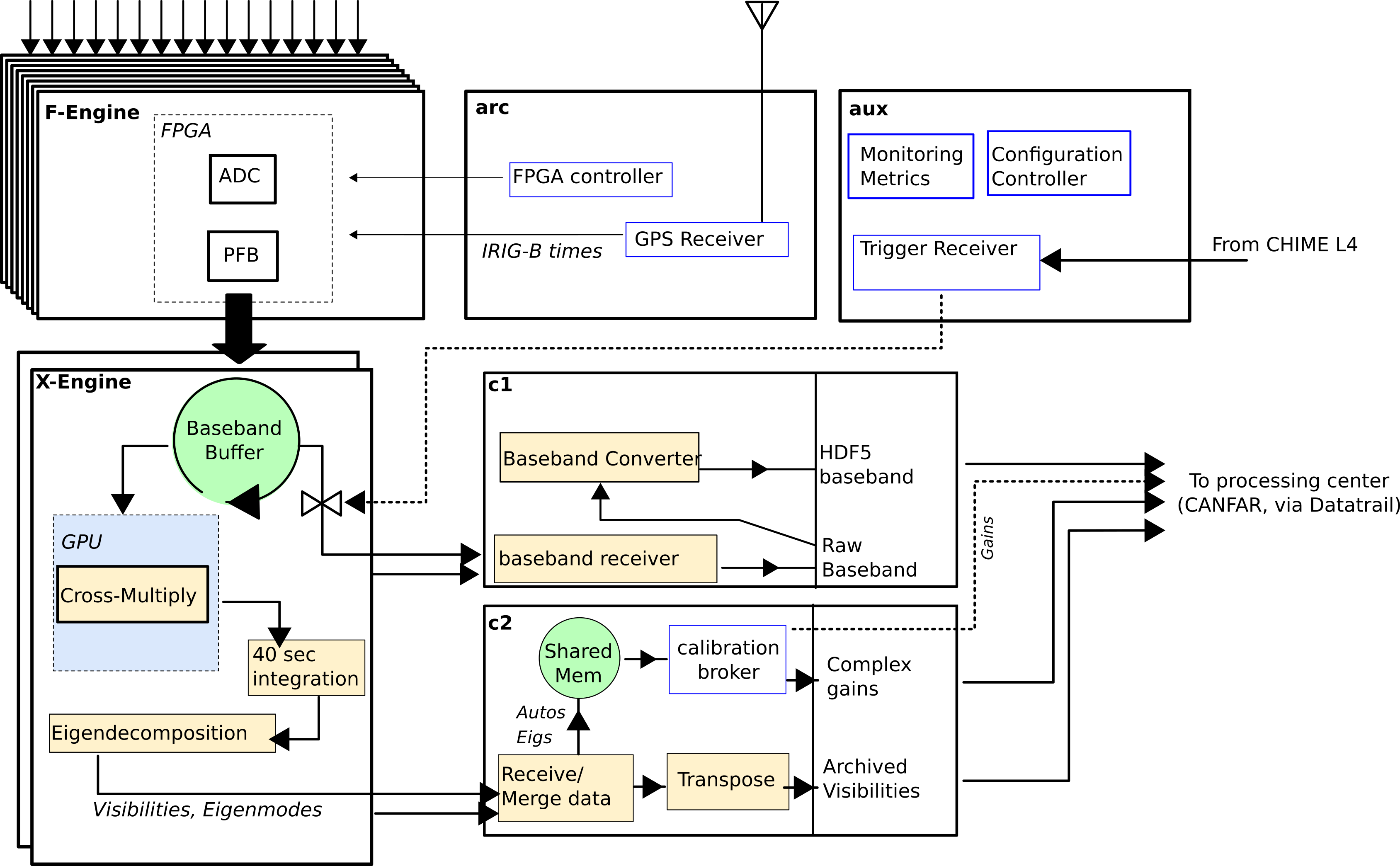}
    \caption{Digital system configuration at KKO, picking up from where Figure~\ref{fig:analog_chain} ended. Black boxes indicate separate computing nodes, with the node name in the corner. Amplified antenna voltages are sampled and channelized in the F-engine into baseband data, which is buffered and correlated in the X-engine. The frequency standard from the Rubidium clock (``Rb clock'') is also sampled in the F-engine via one of the inputs. Visibility data are sent to c2 where they are used for gain calculation, and baseband data are sent to c1 where they are transposed and saved on triggers. The arc and aux nodes handle system management tasks, as well as forwarding triggers from CHIME.}
    \label{f:pipeline}
\end{figure}

\subsubsection{$N^2$ pipeline and gain calibration}
\label{s:N2}

KKO runs a subset of the realtime functionality used to calculate and process the $N^2$ visibilities as CHIME, which is described in Sections~2.4 and~2.5 of the CHIME Overview. The key differences with CHIME are that (1) KKO integrates visibilities over 40~s, instead of 10~s, and (2) visibilities are saved for all baselines, intead of being compressed over redundant baseline groups. The $N^2$ data are used to calculate the complex gain of each antenna feed, and to perform checks of system health and data quality. As in CHIME, this processing is run on a combination of the X-engine and a data receiving node (c2 in this case).

Each GPU node in the X-engine handles half of the frequencies -- calculating the full $N^2$ visibilities, accumulating them to 40~s integration times. This increased integration (relative to CHIME's 10~s) is meant to reduce thermal noise, accounting for KKO having a sixteenth the collecting area of CHIME. The 10~s integration time at CHIME corresponds with the maximum resolution in the East/West direction due to baselines formed between cylinders. Longer integration time would require fringestopping. At KKO (and the other outriggers), the lack of East/West baselines leaves us free to increase the integration time.

The full integrated $N^2$ visibility matrix is passed to a process that iteratively calculates the first four eigenmodes (see \cite{renard_kotekan_2021, wen_accelerating_2017}). The $N^2$ matrix and eigenmodes are then sent to a receiver node, where data streams from all frequencies are merged into one and the full set of visibilities are written to disk. Simultaneously, the eigenmodes and autocorrelation data are put into a shared memory buffer. A Python-based \emph{calibration broker} service maintains a schedule of bright source (``calibrator'') transits and reads data from the shared memory buffer when they occur. During these calibrator transits, the visibility matrix becomes dominated by the first two eigenmodes. By applying fringestop phases to the calibrator position and dividing by the known flux, the eigenvectors yield complex gains which correctly scale input flux to Jansky units and correct per-input instrumental phases (see CHIME Overview, Section~2.5). To remove RFI in the transit data, a mask for persistent RFI is first applied to remove frequency bins that are known to be bad (see Section~\ref{s:flagging}), and then further contamination is identified by comparing on- and off-source eigenvalue ratios to look for statistical outliers. Gain solutions for flagged frequencies are obtained by performing a Gaussian Process regression from good frequencies, using a GPR kernel with 30~MHz width.

\subsubsection{Baseband system}
\label{s:baseband}

Our most critical KKO data products are baseband recordings of FRBs, which can be beamformed, fringestopped, and cross-correlated to perform VLBI with CHIME. These are collected via a triggered baseband recording system, as initially outlined in \citet{abb+18}, but extended to operate over the internet. Baseband data arrive in the X-engine via the fiber optic links to the F-engine. After some initial unpacking, baseband data are placed in a ring buffer, which has the dual purpose of being an intermediary between the network interface controllers (NICs) and the GPUs (see Section~\ref{s:N2}) and to buffer the data. This buffer uses roughly 1~TB of memory on each X-engine node, enabling 37.2~s of baseband data to be held at once. Compared with the 34.2~s buffer at CHIME, this leaves three seconds to accommodate additional latencies between the trigger generation and outrigger baseband writeout.

Outrigger baseband triggers are generated in the L4 system of CHIME/FRB, immediately after the baseband readout stage. When an event meets a minimum S/N threshold, the system packages the event timestamp, baseband readout duration (based on the full DM sweep), the DM, and the DM error into a \emph{payload}. These parameters are estimated by the real-time detection pipeline. Ultimately, only a $\sim 100$~ms frequency-dependent (dispersed) slice of data around the burst will be kept, so the DM is required to perform this selection. The payload is directed to a secondary server, the \emph{dispatcher}, which registers the trigger and holds it to be retrieved by the receiver service at KKO. This portion of the triggering system is managed through the Workflow pipeline management software, in-house software developed to provide a well-defined framework for managing and monitoring CHIME/FRB pipelines and sharing information across networks.

A receiver process on the aux node at KKO checks once per second for new triggers registered in the Workflow system. Once the receiver fetches a trigger, the payload is relayed through a REST request to kotekan on each X-engine node. Kotekan uses the trigger data to compute a per-frequency start time and duration, subsequently processing each frequency independently. The initial memory-to-memory copy out of the primary baseband buffer is very fast (typically about 10\,ms for a 100\,ms dump) compared to the time it takes to write the data to disk on a receiver node, which can create a bottleneck. For this reason, the target data are first copied to a local staging buffer designed to hold multiple baseband dumps as they are transmitted over the network to the node which writes the data to disk. The staging buffer is structured as a bipartite circular buffer \citep{cooke_bip_2003}, with an effective length of 1.7~s.

From the staging buffer, baseband data are moved over the network link to a \textit{baseband receiver} process running on c1 (see Figure~\ref{f:pipeline}). This uses the output data NIC on the X-engine nodes, with 25~G connectivity for the full path through to c1. The baseband receiver is another \texttt{kotekan} process that, upon receipt of the data over the network, writes it in a raw binary format to a local solid state drive (SSD). Each spectral frequency is written independently to its own file. This network transmission and disk write typically takes 5 to 10 minutes for a 100\,ms dump. The final processing stage is the \textit{baseband converter}, a service that converts data from its raw format to our archival \texttt{hdf5} format.

\subsubsection{Management and Monitoring}

Most services running at KKO are run in Docker containers \citep{merkel2014docker}, managed by a Docker Swarm system (in this context, each container is called a ``docker service''). This setup offers a few advantages. In particular, software needs only to be installed in a portable Docker \emph{image}, which can be run at any site without rebuilding code (assuming compatible hardware). This will help to ensure consistency among sites as the other outriggers are constructed. The Docker swarm also runs services to monitor system health, report resource usage, and restart services after disruptions.

Monitoring the conditions at the KKO site is especially important since it is is more remote and less accessible than DRAO. We use an NTI Enviromux-5D to collect data from temperature, humidity, flood, and smoke sensors within the RF-tight room, and a Davis Vantage Pro 2 weather station to monitor temperature, humidity, rainfall, and wind speed and direction outside the hut. All of these metrics, along with data from running services and digital system monitors, are collected using Prometheus\footnote{\url{https://prometheus.io/}} and forwarded to a Grafana\footnote{\url{https://grafana.com/}} server for visualization. An organized schedule of shifts are taken by CHIME/FRB personnel to monitor the conditions at site. The Grafana page is also integrated with a Prometheus Alertmanager\footnote{\url{https://prometheus.io/docs/alerting/latest/alertmanager/}} to send automatic alerts on various triggers. The system is designed to shut down automatically when RF room temperatures are too high, outdoor temperatures are too low (posing a risk to the air conditioner), or on the detection of fire or flood.

The KKO digital system has proved its stability and robustness throughout its commissioning phase. Occasional power outages occurred, and each time the system was successfully brought back to full operation within a few hours of power restoration, typically within an hour. More often, the system was shut down manually out of excess caution during periods of extreme temperature, or because CHIME was offline. The $N^2$ system recorded data for 85\% of the year (between 2022 August 30 and 2023 August 30) and completed over 1000 triggered baseband dumps (mostly scheduled pulsar transits).

\subsubsection{Data management and transport}
\label{s:datatrail}

VLBI cross-correlation, and other data analysis tasks, are handled at a central processing site in the Canadian Advanced Network for Astronomical Research (CANFAR), where CHIME/FRB has sizeable storage and processing resources available. Data products at each site, including the full baseband dumps, are generally small enough that to be transported efficiently over the internet. This transportation is managed by a service called \emph{Datatrail},\footnote{\url{https://chimefrb.github.io/datatrail/}} an automated registration service developed by our team, which logs data and their metadata in a central database, and coordinates transfer operations. Datatrail assigns specific policies for each type of data product, which dictate the conditions for its replication and the conditions under which they can be deleted.

\section{Characterization and Performance: Stand-Alone Telescope}
\label{s:standalone}

\begin{figure}
\centering
	\includegraphics{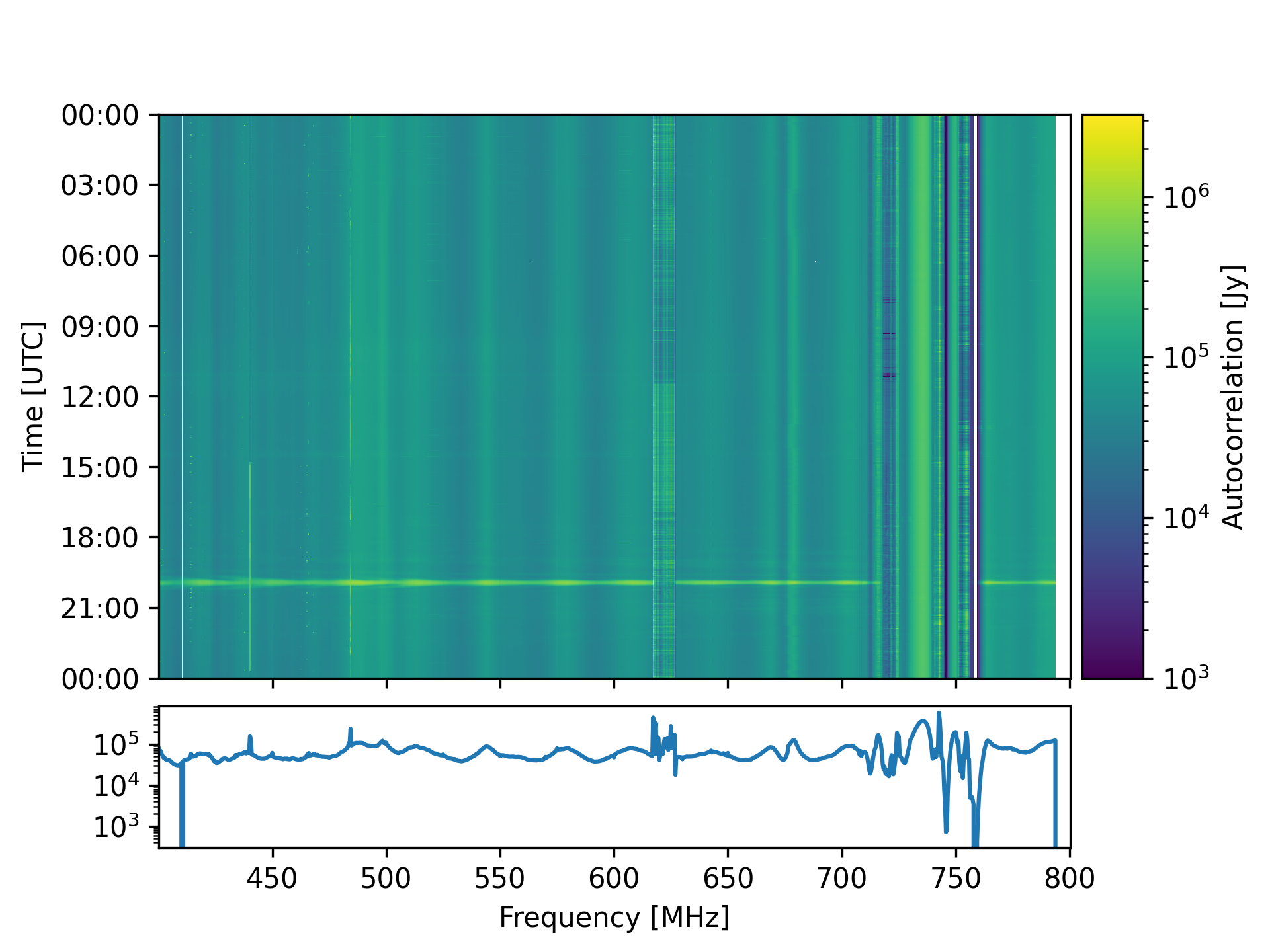}
    \caption{\label{f:waterfall} A waterfall plot showing autocorrelation for a single input over 24 hours. The solar transit can be clearly seen at 20:04 UTC. The lower panel shows the mean over time. The LTE band is distinguishable between 710 and 760~MHz, with a vertical stripe of missing data where the gain calculation failed to converge. No masking has been applied here, but the LTE band, the region around 630~MHz, and a few narrow lower-frequency channels, are all removed from analysis using a static frequency mask (see Section~\ref{s:flagging}).}
\end{figure}

We collected $N^2$ data continuously during the commissioning phase, as well as daily 100-ms baseband dumps of the Crab pulsar. With these datasets, we set out to characterize the system noise, RFI environment, and correlation strengths typical of KKO. Figure~\ref{f:waterfall} shows the autocorrelation in Jy of a single feed over 24 hours on 2023 June 3. The solar transit is clearly visible at 20:04 UTC. Also visible is a ripple with a wavelength of about 30~MHz across the band, which is known from CHIME and arises from reflections between the cylinder and focal line, which amplify the autocorrelation at certain frequencies. There is contamination from LTE above 720~MHz, and a few vertical white stripes at frequencies where the gain calibration failed to converge. For the most part, though, the data appear clean and smooth.

\subsection{Primary beam}
\label{s:primary_beam}

\begin{figure}
\centering
\includegraphics{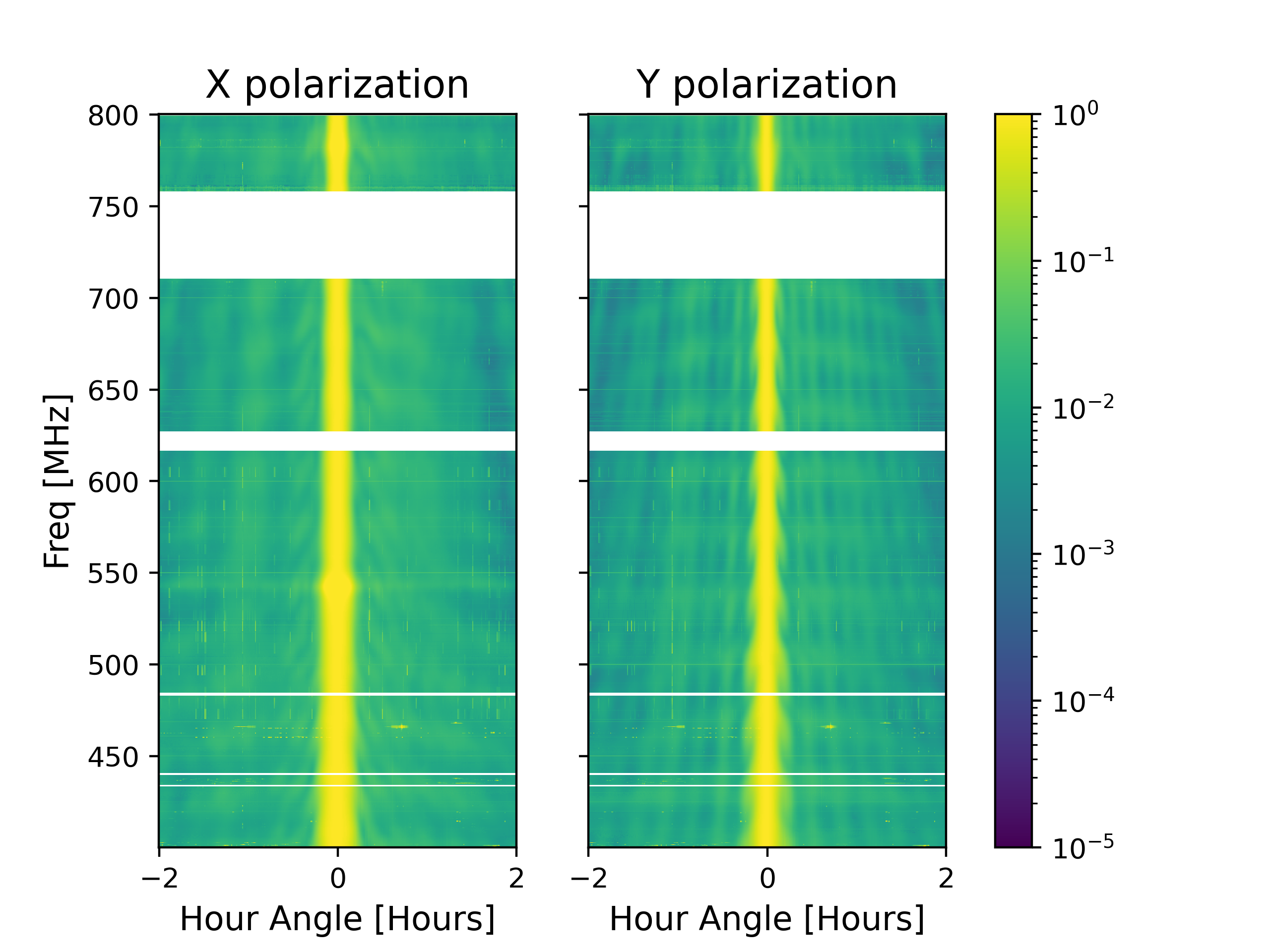}
\caption{Using the solar transit on 2023 June 6 to map the primary beam vs. frequency. The values depicted are the mean autocorrelation (mean taken over input), normalized to the values at 0 hour angle. The hour angle reported is with respect to the CHIME meridian, since the KKO dish is rolled and rotated so that its axis aligns with CHIME's.}
\label{f:solar_beam}
\end{figure}

We looked at the autocorrelation data during a solar transit on 2023 June 6 to map the hour angle and frequency structure of the primary beam. At the time, the Sun was at a declination of $22.6^\circ$. Figure~\ref{f:solar_beam} shows the mean autocorrelation vs frequency and hour angle, with the mean taken over feeds and values normalized to that of zero hour angle. Hour angles are calculated with respect to the CHIME meridian, since the KKO cylinder is rotated such that its beam center aligns with CHIME's. White boxes mark the persistently-contaminated RFI channels (flagging process discussed in Section~\ref{s:flagging}).

We can see from the symmetry about HA = 0 that the cylinder rotation and roll correctly aligned the primary beam with CHIME's. Much of the structure shown in Figure~\ref{f:solar_beam} resembles that seen in Figure~3 of \cite{amiri2022using} -- the main lobe of the Y polarization is narrower than the X polarization, and with sharper sidelobes, and there is a 30-MHz ripple across frequency.

\subsection{RFI Environment and Flagging}
\label{s:flagging}

As mentioned in Section~\ref{s:site}, a new LTE/5G antenna was installed on a mountain near KKO during the time it was being constructed. As a result, the main LTE bands between 700 and 760 MHz are largely unusable, and a custom stop-band filter was installed to suppress power between 718 and 758~MHz.

Aside from the LTE contamination, however, the KKO site is fairly quiet. As a first step toward RFI mitigation, we looked at the statistics of visibility eigenvectors over a 24 hour period to calculate a ``bad channel mask''. Strong RFI will tend to dominate the visibility matrix when there are no bright sources in the field, such that low-rank approximations to the visibility matrix will show mostly RFI. By looking for frequency bins that are statistically distinct, we can identify frequencies that are most often contaminated. We constructed a frequency mask via the following procedure:
\begin{itemize}
	\item Take the 128-element vector $\mathbf{a}(t, \nu) = \sum_i \lambda_i \mathbf{e}_i \mathbf{e}_i^*$ as an estimate of the response of each input, where $\lambda_i$ are the eigenvalues and $\mathbf{e}_i$ the eigenvectors of the visibility matrix, with $i = 0 \ldots 3$.
	\item Take the median absolute deviation (MAD) over time for $\mathbf{a}$ and sum over component:
	\begin{equation}
		N(\nu) = \sum_\text{inputs} \text{median}(|\mathbf{a}(t, \nu) - \tilde{\mathbf{a}}(\nu)|)_t
	\end{equation}
	Here $\tilde{\mathbf{a}}$ indicates the median over time.
	\item Calculate a threshold: $\text{thresh} = \text{mean}(\mathbf{a}) + 4~ \text{std}(\mathbf{a})$, where the mean and standard deviation (std) are taken over all axes.
	\item Mask frequencies for which $N(\nu) > \text{thresh}$.
\end{itemize}

The threshold chosen for identifying outliers is fairly conservative -- 4$\sigma$ over the mean amplitude in the instrument, assuming Gaussian noise at most times. This threshold is biased upwards by the presence of bright source transits in the data, which will drive the mean higher across all frequencies. The MAD is less sensitive to outliers than the standard deviation.

With this procedure, we can identify most of the LTE-contaminated channels, as well as several digital TV channels between 425 and 480~MHz and another contaminated band around 625~MHz. We supplement this with the stop-band of the LTE notch filter to construct the final static RFI mask. Altogether, this mask removes 15\% of frequency channels from the data. This mask is meant to identify frequencies that are consistently filled with RFI, and is only recomputed if we see new contaminants arise. So far, the mask computed from 2023 June 6 data has been sufficient.

\subsection{Calibration stability}
\label{s:stability}

\begin{figure}
	\centering
	\hspace*{-.1\linewidth}
	\includegraphics[width=1.2\linewidth]{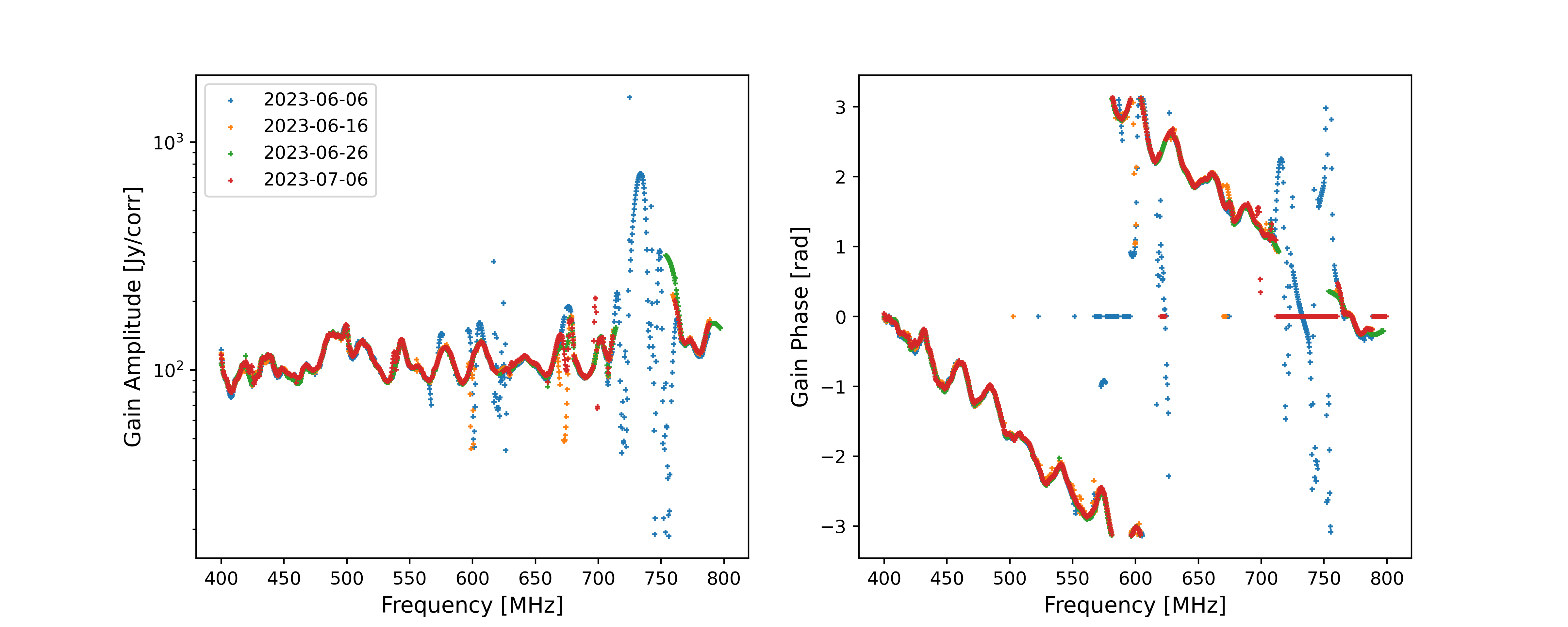}
    \caption{Amplitude and phase of the gain for a single input on several days. These are all computed from Cygnus A transits. The persistent RFI mask (see Section~\ref{s:flagging}) was applied to the data starting on 2023 June 7, with the effect of removing some of the noisier structure visible on 2023 June 6. Gains are largely stable over many weeks.}
   	\label{f:calibration}
\end{figure}

The complex daily gains computed from the $N^2$ data stream are very stable. Figure~\ref{f:calibration} shows the amplitudes and phases of a single input's complex gains calculated from Cygnus A transits over a month. Between the first plotted day (2023 June 6) and the second, the persistent RFI mask described in Section~\ref{s:flagging} was implemented, which cut the noisier RFI-contaminated channels from the analysis, and hence improved the overall gain solution. Figure~\ref{f:calibration} also clearly shows the 30-MHz ripple from reflections

\subsection{Noise Characteristics}
\label{s:noise}
\subsubsection{SEFD Characterization}
The system equivalent flux density (SEFD) is the flux of a hypothetical point source which would double the system temperature when at the primary beam center \citep{thompson_interferometry_2017, ww99}. SEFD characterizes the thermal noise of the system in a way that can be compared with observed sources. To estimate the SEFD of each feed, we examined the autocorrelations over a short period of time when the system was stable and no bright sources were overhead -- between 08:00 and 08:30 UTC on 2023 August 30. 

\begin{figure}
	\centering
	\includegraphics[width=\linewidth]{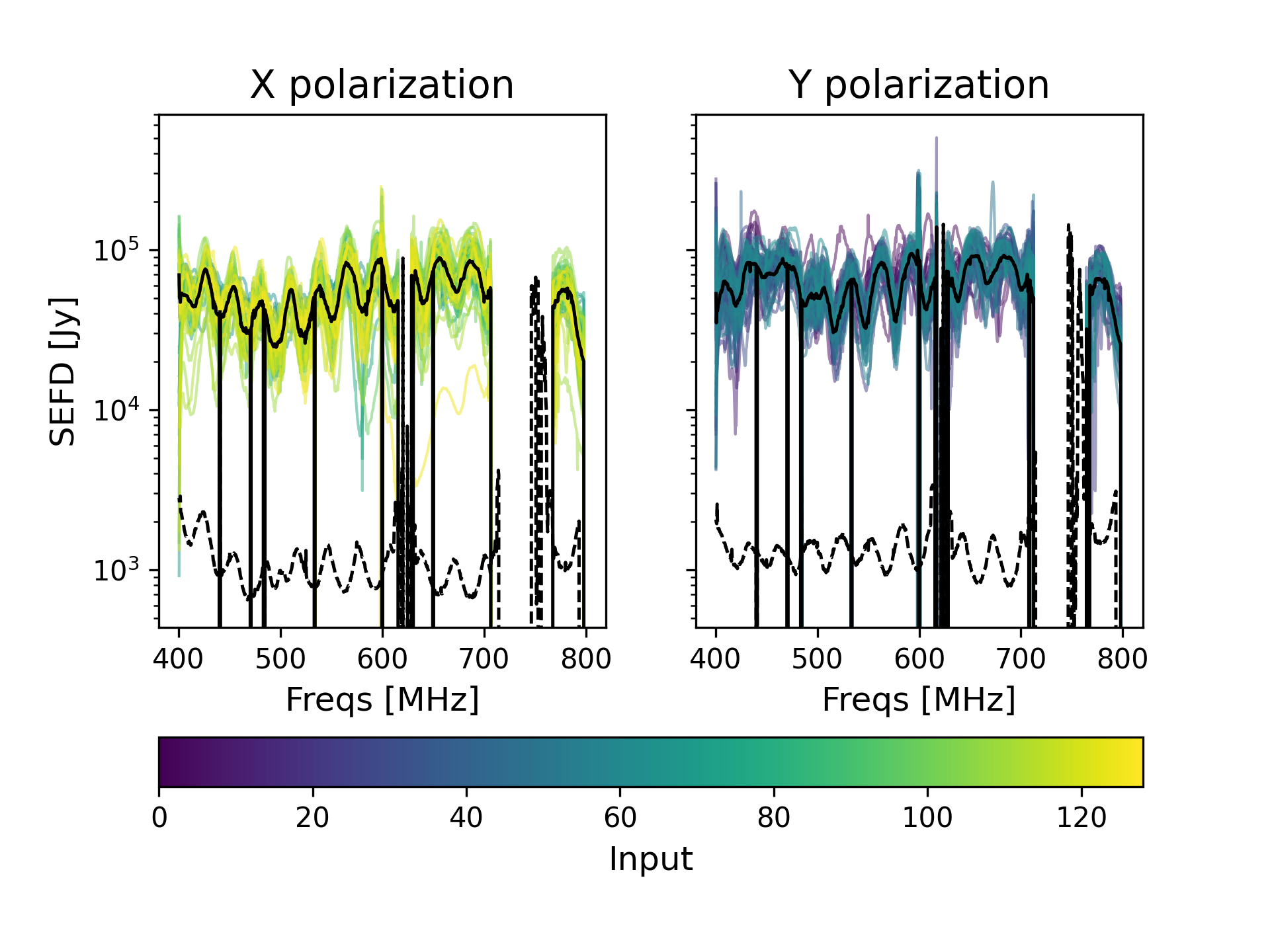}
    \caption{System equivalent flux density (SEFD) of each input. Color indicates a specific input number. Inputs 0 to 63 are all X polarization, and 64 to 127 are Y pol. The solid black line shows the median over inputs within each polarization. The dashed black line marks the SEFD of beamformed baseband data.}
    \label{f:sefd}
\end{figure}

We extracted the autocorrelations from the $N^2$ visibility data saved for that period and calibrated them by applying the gains calculated from the Cas~A transit about seven hours later. Taking the mean over time, we get an estimate of SEFD for each individual input. Figure~\ref{f:sefd} shows the SEFD vs. frequency for each KKO input, split by polarization, as well as the median over all inputs. The scale of $\sim 4 \times 10^4$~Jy corresponds to a system temperature of about 113~K, assuming an effective area of 7.8~m$^2$ for each feed. These observed values are consistent with the equivalent results from CHIME, as expected since the analog hardware is the same. The dips observed around 425~MHz, 620~MHz, and above 700~MHz are all due to known RFI contamination skewing the measured gains.

We also plot the SEFD of beamformed baseband data in Figure~\ref{f:sefd} as a black dashed curve. To obtain these values, we take a 500~ms sample of baseband data during a transit of Cyg A and beamform it to two positions -- $P_\text{on}(\nu)$ is the power after beamforming toward the position of Cyg A, and $P_\text{off}(\nu)$ is the power after beamforming 20$^\circ$ north of Cyg A, in a relatively dark patch of sky. The beamformed SEFD is then obtained by taking
\begin{equation}
	\text{SEFD}_\text{bf}(\nu) = P_\text{off}(\nu) \frac{F(\nu)}{P_\text{on}(\nu) - P_\text{off}(\nu)}
\end{equation}
where $F(\nu)$ is the known flux density of Cyg A. The fraction gives a rough estimate of the gain, scaling the off-source power into Jansky units. The beamformed SEFD values should be lower than those of individual elements because the SEFD is inversely proportional to the effective area of the instrument. The effective area of the beamformed data is equivalent to the full array area, a factor of about 1000$\times$ that of a single feed.

\subsubsection{Radiometer Test}

Astrophysical radio sources will induce voltages in the receiving antenna which are independent and identically distributed Gaussian variables. As a result, the autocorrelation power $P$ is $\chi^2$-distributed with one degree of freedom, and the ratio of the mean power to the variance is
\begin{equation}
    R \equiv \frac{\langle P \rangle }{\sqrt{\text{Var}[P]}}\frac{1}{\sqrt{\tau\Delta\nu}} = 1.
    \label{eqn:radiometer}
\end{equation}
This is the \emph{radiometer equation} \citep[see, e.g.][]{thompson_interferometry_2017}. The factor of $\sqrt{\tau \Delta \nu}$ comes about from averaging $N = 2 \tau \Delta \nu$ samples when integrating over a time $\tau$. If the autocorrelations in the system deviate from this relation, it indicates that a non-Gaussian signal is present, which can be caused by malfunctioning components.

We compute the quantity $R$ defined in Equation~\ref{eqn:radiometer} for three integration timescales: 1~ms, 10~ms and 100~ms. These integration timescales are well-motivated --- CHIME performs its FRB search with a 1~ms integration window, while some FRBs have total durations (including dispersion measuring and scatter-broadening) around 10~ms and in some cases even 100 ms. Furthermore, continuum sources used for VLBI calibration (see Section~{\ref{s:vlbi_processing}}) are integrated on the order of 100-ms timescales. Using a 1400-ms full-array baseband dump obtained during June of 2023, we beamform KKO to a quiet location on the sky, then integrate the power over each integration timescale (resulting in $n = 1400,~140~\text{and}~14$ integrations, respectively). For each frequency bin, we compute the mean and variance over an ensemble comprising all feeds within a given polarization and $n=14$ integrations from the full 1400-ms duration (this limitation is chosen to ensure the same ensemble size for all three integration lengths). We do not perform any frequency averaging and retain our native frequency resolution of $\Delta\nu=390.625$~kHz.

\begin{figure}
    \centering
    \includegraphics[width=1\textwidth]{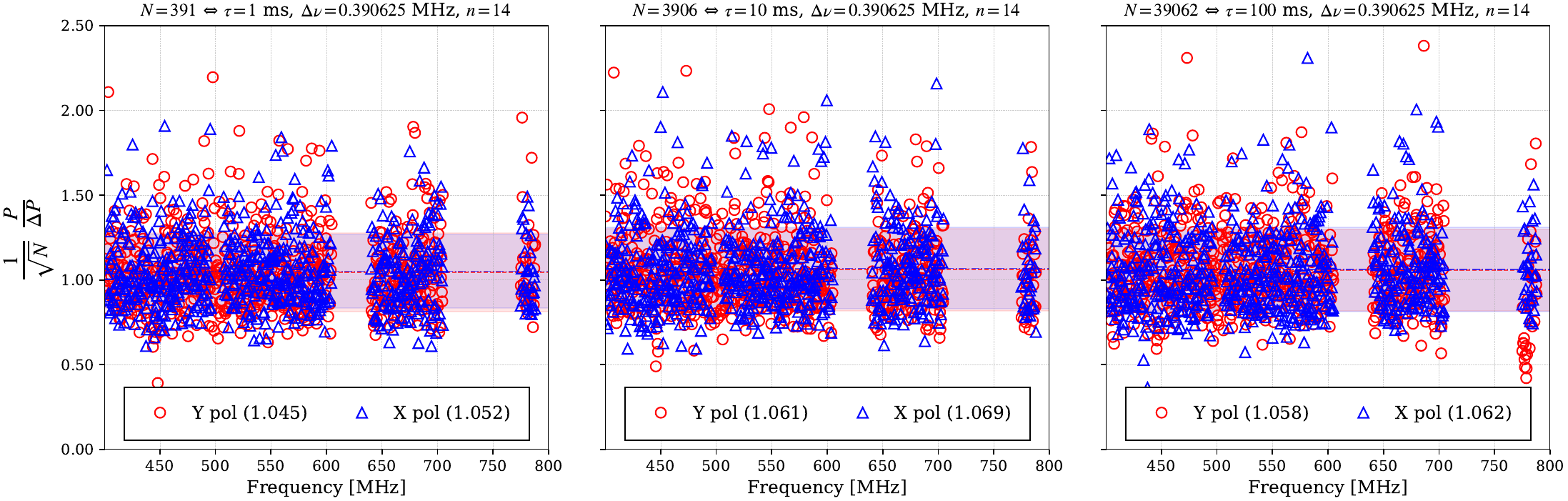}
    \caption{Radiometer test on 1, 10 and 100 ms integration timescales with KKO full-array beamformed data. Using a 1400 ms full-array baseband dump, we integrate the power in each frequency bin over each timescale, after removing RFI-contaminated frequencies. For an equivalent comparison among the three plots, we compute the average and variance using $n = 14$ independent measurements and normalize the ratio by the square root of the number of samples. We perform the test independently between Y (open red circles) and X (open blue triangles) polarizations. The median of the ratio for each polarization is provided in the legend. $\pm1\sigma$ regions are indicated by the highlighted horizontal regions. }
    \label{fig:radiometer}
\end{figure}

Figure~\ref{fig:radiometer} shows the quantity $R$ in Equation~\ref{eqn:radiometer} for all frequencies and all three integration times. Marker color and shape indicate feed polarization --- Y polarization is represented by red circles, and X polarization by open blue triangles. The median of the ratio for each polarization is shown in the legend (max $\sim 1.069$, min $\sim 1.045$). Red and blue horizontal highlighted regions represent the $\pm1\sigma$ region of the Y and X polarization data points, respectively. We perform aggressive RFI flagging and remove frequency channels known to perform poorly, as these points end up at very large values and obscure the behavior in cleaner parts of the band. For all three integration windows, the median of the ratios fall within $1\sigma$ of the expected value of 1. Each median tends to skew slightly above 1, due to the small number of samples in each frequency bin. We chose to restrict this plot to 14 samples per frequency bin since this is the maximum samples available on the longest integration time, and using the same for all three made for an easier comparison. Increasing the samples per frequency bin on the shorter integration times, the median drops down closer to 1.

We perform a similar test on a daily basis to monitor the health of the tied-array baseband data using a 100-ms baseband dump. This daily test should help us identify non-thermal systematics in our data should they arise.

\subsection{Performance}

\subsubsection{Interferometry}
\label{s:interferometry}

To assess the performance of KKO as a standalone radio interferometer, we perform two independent tests. First, we produce an all-sky map as seen by KKO by beamforming $N^2$ data to a range of declinations at each time sample over 24 hours (a \emph{ring map}). Although imaging with a compact array with low UV coverage like KKO will not produce much detail, it can verify that bright sources appear in the correct positions, and that our computed complex gains are correct. The second test is to compute the feed positions of each polarization pair using known, radio-bright calibrators. This more precisely verifies that our understanding of the array geometry is correct.

\begin{figure}
    \centering
    \includegraphics[width=1\textwidth]{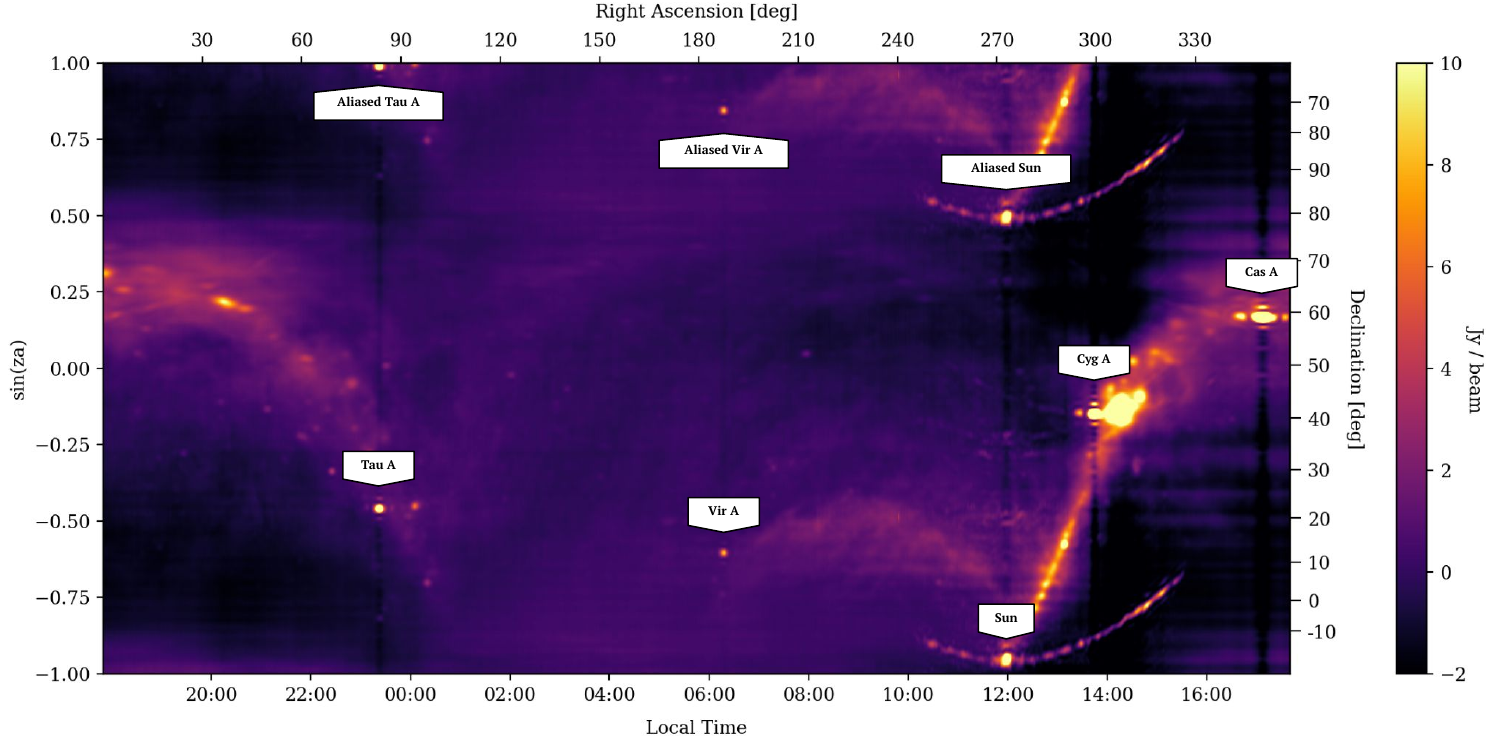}
    \caption{Map of the northern radio sky at 679~MHz constructed from data collected by KKO over a single sidereal day (2022 December 23 to 24). YY $N^2$ visibility data (excluding autocorrelations) were beamformed to a grid of 2048 declinations along the meridian, spanning from horizon to horizon, equally spaced in $\sin($ZA$)$. The median of each set of autocorrelation data was subtracted in order to reduce the effects of feed crosstalk. The Sun and the four brightest point sources are identified, along with their aliased images. The map is shown with local time increasing from left to right. Compare with the CHIME equivalent plot in Figure~28 of the CHIME Overview paper.}
    \label{fig:ringmap}
\end{figure}

Figure~\ref{fig:ringmap} shows a ring map of the Northern sky at 679~MHz, using data collected by KKO over a single sidereal day (2022 December 23 and 24). We construct this ring map using the methods described in Section~4.5.1 of the CHIME Overview: For every 40-s time step, we calibrate the data using complex gain solutions calculated during the Cyg A transit on December 23, then beamform YY $N^2$ visibility data to a grid of 2048 declinations, equally spaced in sine of zenith angle (ZA), along KKO's meridian. The horizontal axis is given in local time (Pacific Standard Time, UTC - 8) with R.A. and decl. plotted along the top and right, respectively.

Comparing Figure \ref{fig:ringmap} to Figure~28 in the CHIME Overview, many similarities and some notable differences emerge. First, as with the CHIME ring map, radio-bright sources (our Milky Way galaxy, the Sun, Cyg A, Cas A, Tau A and Vir A) are all clearly visible. Second, aliased versions of the Sun, Tau A and Vir A are also visible, although the aliased antipodal versions of Cas A and Cyg A are not. The characteristic ``smile'' sidelobe structure of bright sources in CHIME's ring map are either no longer noticeable (for Tau A, Vir A) or significantly dimmer (for the Sun, Cyg A, Cas A) in KKO's ring map. These differences can be attributed to KKO's reduced sensitivity and resolution: with 1/16 of the total collecting area of CHIME, dimmer point sources are no longer visible; the overall map is blurrier. The sidelobe of the Sun is cut off at Local Time < 10:00 due to Copper Mountain blocking the south-eastern sky in the early morning.

As a check of the robustness of the phasing, we reconstruct the feed positions from $N^2$ data. As in Section~\ref{s:N2}, we assume that when a bright source $A$ at position $\hat{s}_A$ is in the field, the visibility matrix may be expressed as a low-rank approximation,
\begin{equation}
	\mat{V} = \sum_a \mathbf{v}_a \mathbf{v}_a^\dagger,
\end{equation}
where $a= 0,\ldots,3$ indexes eigenmode. Each eigenvector component is
\begin{equation}
	v_j(t, \nu) = g_j \sqrt{T_A} \exp\left(  -2\pi i (\nu/c) \mathbf{r}_j \cdot \hat{s}_A(t) + \phi_j + \phi_0 \right),
	\label{eqn:eigenmode}
\end{equation}
where $\phi_0$ is an arbitrary reference phase common to all feeds, $g_j$ is the gain amplitude, $\phi_j$ is a feed-specific instrumental phase, and $\mathbf{r}_j$ is the $j^\text{th}$ feed position relative to some chosen origin. $T_A$ is the brightness of the source, weighted by the primary beam function. Since the feed positions appears in the fringe term of Equation~\ref{eqn:eigenmode}, we ignore the visibility amplitude and focus only on the phase.

Letting $\mathbf{r}_j = (x_j, y_j, z_j)$, where each component corresponds to East, North, and Up directions, the source position $\hat{s}_A$ in this frame is related to the hour angle (HA) and declination ($\delta$) of the source via
\begin{equation}
	\hat{s}_A = \begin{pmatrix}
		-\cos\delta \sin\text{HA}\\
		\cos l \sin\delta - \sin l \cos\delta \cos\text{HA}\\
		\sin l \sin\delta + \cos l \cos\delta \cos\text{HA}
	\end{pmatrix},
\label{eqn:topo_pos}
\end{equation}
where top to bottom are $(X, Y, Z)$ components, and $l$ is the latitude of the observatory.

At transit, HA$=0$, so the X component vanishes. Further, $\sin \text{ZA} = \cos l \sin\delta - \sin l \cos\delta$. Taking $z=0$ for all feeds, the result is that the eigenvector phase during a source transit is:
\begin{equation}
	v_j(t_A, \nu) = \exp \left( -2\pi i y_j \sin\text{ZA}_A \nu / c  + \phi_j \right),
\end{equation}
where $t_A$ is the transit time of source $A$, and ZA$_A$ is the zenith angle of the source at transit. To remove the instrumental phase term, we divide this eigenmode by that of a separate bright source transit (source $B$), which gives us a \emph{phase-referenced eigenvector}:
\begin{equation}
	\tilde{v}_j(t_A, t_B, \nu) = \exp \left( -2\pi i y_j (\sin\text{ZA}_A -\sin\text{ZA}_B) \nu / c \right).
\end{equation}
This form assumes (reasonably) that the instrument phase is constant between the transits of $A$ and $B$. The Fourier conjugate of the observing frequency $\nu$ is a quantity proportional to $y_j$, and so we can calculate the North/South feed position by finding the peak of the Fourier transform.

To perform the analysis, we take a Cyg A transit as our source $A$ and a Cas A transit from the same day as source $B$. We sum over the first four eigenmodes for each transit, then divide the summed eigenvectors. After flagging out the RFI-contaminated frequencies using the persistent RFI mask (Section~\ref{s:flagging}), we use a Lomb-Scargle periodogram to estimate the Fourier transform, since it is more robust to gaps than a standard FFT \citep{lomb1976least, scargle1982studies}.

\begin{figure}
	\includegraphics[width=\linewidth]{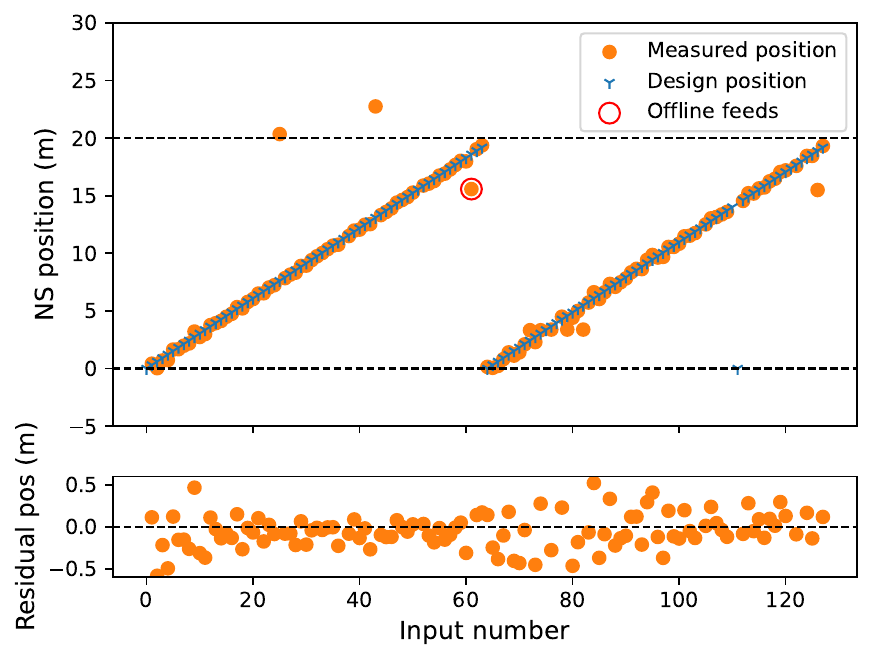}
    \caption{North-south position of the dual polarization feeds. The top panel shows the north-south position of the feeds as a function of feed number. The first 64 feeds are the Y-polarization feeds and the next 64 feeds are the X-polarization feeds. The bottom panel shows the residual when the measured position is compared to the design position.}%
    \label{f:feedpositions}
\end{figure}

Finding the peaks of each transform, and scaling the peak location accordingly, yields the feed positions shown in Figure~\ref{f:feedpositions}, relative to feed 0 (the southernmost) for the Y-polarization feeds and feed 70 (a chosen good feed) for the X-polarization feeds. Except for a handful of feeds that were deactivated in the data analyzed, the positions are all generally within a few centimeters of their design position.

Although it is theoretically possible to measure the East (X) component of the feed positions, we do not do so here because the expected positions are less than 10~cm. The uncertainty in this analysis process, which is mostly set by the available baseline lengths, is greater than that.

\subsubsection{Baseband Performance on Pulsars}
\label{s:baseband_performance}
In this section, we characterise KKO's baseband performance on millisecond radio transients by testing that the signal-to-noise ratios of pulsar observations are consistent with expectations. The radiometer equation gives us that, for a single source in the field of view, the signal to noise ratio of $N$ coherently added antenna signals is 
\begin{equation}
	\SNR = C\frac{T_\text{A}}{T_S}\sqrt{\frac{N(N-1)\Delta\nu\tau}{2}},
\end{equation}

where $T_A$ is the antenna temperature due to the source, $T_S$ is the system temperature, $\Delta \nu$ is the frequency channel width, and $\tau$ is the integration time. $C\leq 1$ is a factor that depends on the properties of the receiver. This is true under the assumption that $T_A \ll T_S$ and that noise is uncorrelated among antennas. We compare the observed S/N at KKO to that at CHIME, making the reasonable assumption that $C$ is the same for both due to their identical receiver hardware. The expectation is, thus:
\begin{equation}\label{eq:SNR_ratio}
	\frac{(\SNR)_{\text{KKO}}}{(\SNR)_{\text{CHIME}}} = \sqrt{\frac{N_\text{KKO}(N_\text{KKO} - 1)}{N_\text{CHIME}(N_\text{CHIME} - 1)}} \approx \frac{1}{16} = 0.0625,
\end{equation}
where $N_\text{KKO} = 128$ and $N_\text{CHIME} = 2048$. Comparing single-feed SEFDs between CHIME and KKO has established that they have similar system temperature (see Section~\ref{s:noise}), as expected since they comprise nearly-identical analog hardware and are suspended over similar-area cylinders. We assume, justified by the observed bright source transits, that the CHIME and KKO primary beams overlap sufficiently well that each instrument receives a similar response to incoming signals.

\begin{figure}
    \centering
    \includegraphics[width=1\textwidth]{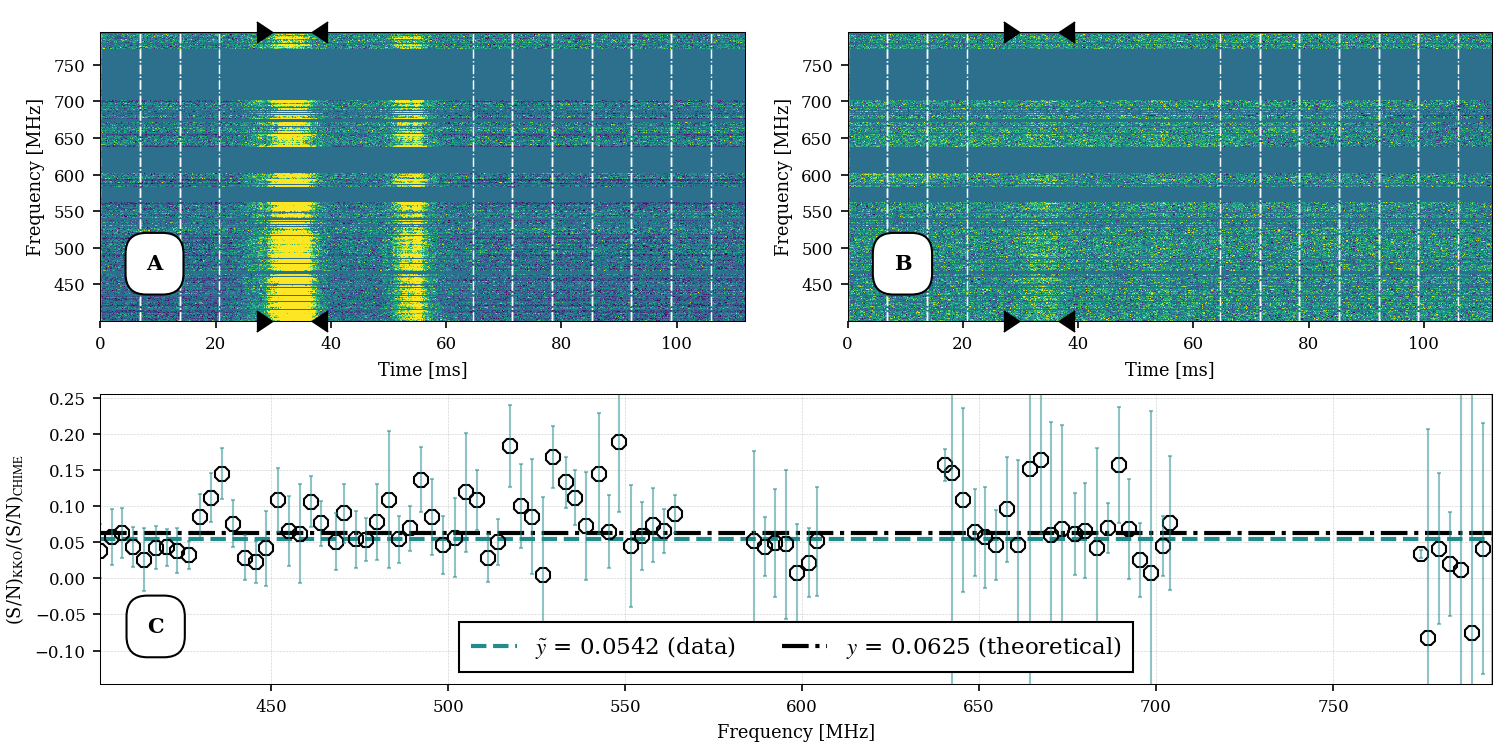}
    \caption{Comparison of baseband S/N of a detected PSR B0329+54 pulse between CHIME and KKO. \textbf{Panel A:} Waterfall plot of a B0329+54 pulse detected by CHIME. The horizontal axis covers the full duration of the baseband dump. Vertical white lines mark off-pulse regions and black carets mark the on-pulse region used to estimate S/N. \textbf{Panel B:} Same as panel A but for KKO. \textbf{Panel C:} Ratio of the per-channel S/N as a function of frequency. Data are plotted with open black circles; uncertainties in teal. The expected value is represented by the black, dash-dotted line, and the median ratio across all (non RFI) frequency channels is shown by a dashed teal line.}
    \label{fig:snr_comparison}
\end{figure}

We evaluate Equation~\ref{eq:SNR_ratio} using a PSR B0329+54 pulse on 2023 June 20. Raw baseband data from both telescopes were beamformed towards a nominal (J2000) position of $(\alpha,\delta) = (03^\mathrm{h} 32^\mathrm{m}59^\mathrm{s}.4096, 54^\circ 34' 43''.329)$ and incoherently dedispersed to a DM of $26.7641$ pc cm$^{-3}$ \citep{hmth04}. The time window over which the signal was integrated was determined by measuring where the integrated intensity profile dropped by 20\% of the peak value. To estimate the noise, nine off-pulse regions with equal width to that of the signal were identified. The per-frequency S/N was estimated by calculating
\begin{equation}
    (\SNR)_\nu = \frac{I_{\nu,\text{ON}} - \bar{I}_{\nu,\text{OFF}}}{\text{MAD}(I_{\nu,\text{OFF}})},
    \label{eq:SNR_per_freq}
\end{equation}
where $I_{\nu,\text{ON}}$ is the per-frequency intensity of the on pulse, $ \bar{I}_{\nu,\text{OFF}}$ is the mean and $\text{MAD}(I_{\nu,\text{OFF}})$ is the median absolute deviation of the off-pulse regions, respectively. The results are given in Figure~\ref{fig:snr_comparison}. Panels A and B are waterfall plots of the B0329+54 pulse detected by CHIME and KKO, respectively, with vertical lines marking the on/off pulse regions used for analysis, and with color ranges scaled to show the relative weakness of the KKO signal compared with CHIME. We have averaged over 78 time samples (corresponding
to 0.2~ms segments) and limited the color range to $\pm 5\sigma$, where $\sigma$ is the standard deviation in the off-pulse regions of CHIME data. Keeping the same color scale for both panels A and B demonstrates the relative weakness of the KKO signal. Frequency channels removed due to RFI have been filled with the median value of the (non-RFI) off-pulse regions.

Panel C of Figure~\ref{fig:snr_comparison} shows the S/N per-frequency channel. Open face points represent the data; uncertainties are in teal. For visualisation purposes, we average over $6~\mathrm{MHz}$ segments and propagate errors by summing per-frequency uncertainties in quadrature. The theoretical value of 0.0625 is indicated by a solid, horizontal black line. The median across all frequency channels is $\tilde{y} = 0.0542$ and is represented by a dashed teal line. Overall, the behavior of the ratio of S/N is expected. The median value suggests that KKO is performing at $\sim 85\%$ of the optimal configuration across the entire bandwidth (after RFI removal). The reason for not achieving a median value of 0.0625 can be explained by imperfect RFI removal, not taking into account flagged feeds (at CHIME and KKO) and the assumption of identical feed response between both polarizations. Regardless, for the purposes of VLBI with KKO, the measured median is sufficient to tell us that the array is functioning as a transient detector.

\section{Characterization and Performance: VLBI}
\label{sec:vlbi}

\subsection{Overview of data processing}
\label{s:vlbi_processing}
The full details of the CHIME/FRB Outriggers VLBI analysis are covered in \cite{leung_andrew_pyfx}. This section gives a brief overview of the technique. To function as a VLBI station, KKO produces raw baseband data dumps, which are calibrated with the daily gain solutions, beamformed to a target position along the cylinder axis, and cross-correlated with similarly calibrated and beamformed CHIME data. This cross-correlation produces a set of visibilities vs. time and frequency, corresponding to a single baseline (between the phase centers of CHIME and KKO).

For each target source, we identify several continuum radio sources in the field of view to use as calibrators and compute cross-correlation visibilities for each. Dividing the target visibility by a calibrator visibility removes direction-independent instrumental phases. We model the phase of this \emph{phase-referenced} visibility as a sum of dispersive and non-dispersive terms:
\begin{equation}
	\label{eq:phase_model}
	\phi(\nu) = (2 \pi \tau) \nu  + (2\pi K) (\Delta\text{TEC}) / \nu.
\end{equation}
In the above, $\tau$ is the relative geometric delay between the target and the calibrator, $\Delta\text{TEC}$ is the difference in ionospheric total electron content between the lines of sight, and $K = 1.345 \times 10^3$~MHz~TECU$^{-1}$ is a constant \citep{Weeren_2016}. Information on source position (relative to the calibrator) is contained in $\tau$.

We solve jointly for $\tau$ and $\Delta \text{TEC}$ by maximizing the posterior probability, which under the assumption of a uniform delay and $\Delta\text{TEC}$ is equivalent to the log-likelihood:
\begin{equation}
	\label{eqn:loglikelihood}
	\log \mathcal{L}(\tau, \Delta \text{TEC}) = \sum_k \frac{|\mathcal{V}_k|}{\sigma_k^2} \; \Re \left[ \mathcal{V}_k \exp\left\lbrace-i \phi(\nu_k)\right\rbrace\right].
\end{equation}
In the above expression, $k$ indexes frequency bins, $\mathcal{V}_k$ denotes the measured visibility in the $k^\text{th}$ frequency bin, and $\sigma_k^2$ is the measured variance of the visibility. The function $\Re ()$ returns the real part of its argument. This likelihood function, derived in \cite{leung_andrew_pyfx}, is based on the assumption that the spectrum of the source is well-approximated by the measured visibility amplitudes, which is a fair assumption when a single bright source dominates the field of view.

\subsection{Sensitivity}
\label{s:vlbi_sensitivity}
To assess the CHIME --- KKO baseline sensitivity, we compare a pulsar's S/N from VLBI cross-correlation with the S/N from an autocorrelation at CHIME. This lets us compare with a theoretical expectation that does not depend on source brightness or aperture efficiency parameters (see Section~\ref{s:baseband_performance}).

In this case, the S/N of the cross-correlated signal is related to the autocorrelation of CHIME by
\begin{equation}
    (\SNR)_\text{cross} = (\SNR)_{\text{CHIME}}\sqrt{\frac{2A_{\text{KKO}}}{A_{\text{CHIME}}}} = \frac{1}{\sqrt{8}}(\SNR)_{\text{CHIME}},
\end{equation}
where $A_{\text{KKO}}$ and $A_{\text{CHIME}}$ are the collecting areas of KKO and CHIME, respectively \citep{mlc+22}. Using the same PSR~B0329+54 baseband data used in Figure~\ref{fig:snr_comparison}, we beamform to the pulsar position and cross-correlate using \texttt{PyFX}, which includes the same windowing function for finding the on-pulse region as was used in the previous analysis. This gives us a set of visibilities vs frequency and integer delay, measured in units of ($2.56$~$\mu$s). Since the data are fringestopped to the known pulsar location, the peak of cross-correlation should (and does) occur at zero delay. The other integer delays, which are equivalent to fringestopping and correlating towards positions separated from the pulsar, are taken to estimate noise.

\begin{figure}
    \centering
    \includegraphics[width=1\textwidth]{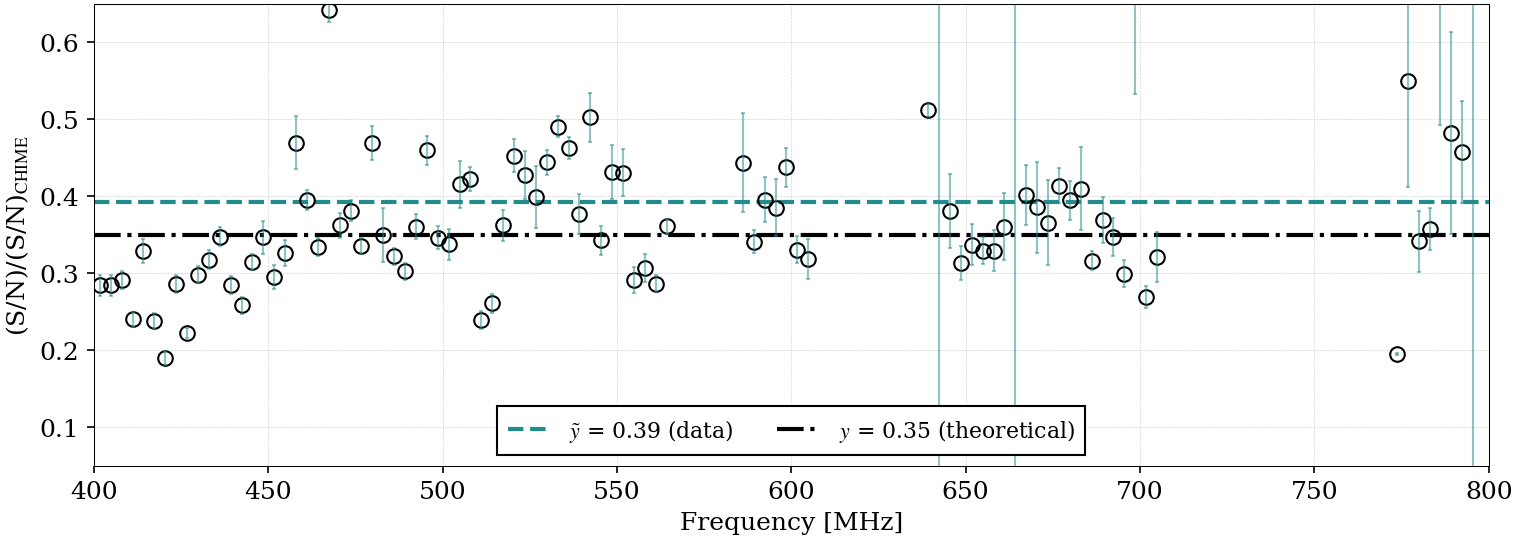}
    \caption{Cross correlation performance on a PSR B0329+54 pulse on the CHIME-KKO baseline. Using the same event to test the baseband performance in Figure \ref{fig:snr_comparison}, we correlate and compute the S/N of both the cross-correlated visibilities and CHIME autocorrelations and plot their ratio. The median value across all clean channels is 0.39 and is indicated by the teal, dashed line. The theoretical ratio of 0.35 is shown by the black, dash-dotted line. For visualization purposes, we average over 6-MHz segments to highlight the performance at lower versus higher frequencies. Error bars are present but are small in comparison to the spread in data points. }
    \label{fig:xcorr_snr}
\end{figure}

To compute $(\SNR)_\text{cross}$, we apply Equation~\ref{eq:SNR_per_freq} with $I_{\nu, \text{ON}}$ being the on-pulse (lag 0) intensity and $I_{\nu, \text{OFF}}$ being the off-pulse data (nonzero integer delay). $(\SNR)_\text{CHIME}$ is computed as in the previous section. Figure~\ref{fig:xcorr_snr} shows $(\SNR)_\text{cross} / (\SNR)_\text{CHIME}$. For easier visualization, we average data over frequency in 6-MHz segments as is done in Figure \ref{fig:snr_comparison}. The median value across all uncontaminated channels prior to averaging is $0.39$, in good agreement with the expectation of $1/\sqrt{8} \approx 0.35$. Although most contaminated channels have been removed, we limit the y-axis further to exclude any channels still contaminated with RFI after averaging. We highlight the drop in performance at lower frequencies which is likely the result of RFI: Figure \ref{fig:snr_comparison} highlights a similar drop in KKO's baseband performance at the lowest frequencies. Performing these combined analyses on multiple PSR B0329+54 pulses reveals that the poor performance at lower frequencies does not persist across all events, suggesting that some time variable signal (RFI) is contaminating the lower end of the band in this particular case.

\subsection{Clock stability}
\label{s:clock_stability}

The clocking system described in Section~\ref{s:clocking} can be used to calculate the deviation $\tau^{\text{GPS}}_{\text{Rb}}$ between the GPS clock and the Rubidium (Rb) clock at KKO, and similarly the deviation $\tau^{\text{GPS}}_{\text{maser}}$ between the GPS clock and the hydrogen maser at CHIME. The cross-correlated visibilities encompass a delay $\tau_{\text{clock}}$ due to these clock deviations:
\begin{equation}\label{eq:tau_clock}
    \tau_{\text{clock}} = \tau^{\text{GPS}}_{\text{maser}} - \tau^{\text{GPS}}_{\text{Rb}}.
\end{equation}
Since the characteristic timescale of timing variations in the GPS module is $\gtrsim$1~s, $\tau_{\text{clock}}$ is stable for $\lesssim$1~s. Thus, if the target and the calibrator are separated by more than 1~s, the phase-referenced visibility will have an additional clock delay $\tilde{\tau}_{\text{clock}}$ which can be calculated using the clock stabilisation pipeline:
\begin{equation}\label{eq:clock_correction}
    \tilde{\tau}_{\text{clock}} = \tau_{\text{clock}}(\text{t}_{\text{target}}) - \tau_{\text{clock}}(\text{t}_{\text{calibrator}})
\end{equation}
where $\text{t}_{\text{target}}$ and $\text{t}_{\text{calibrator}}$ are the start times of the target and calibrator baseband dumps respectively. \\

We test the accuracy of the clock stabilization pipeline, and thus the timing precision of the Rb clock, on a timescale of $\sim$10$^{3}$~s by phase-referencing a sequence of steady source observations to a single time and observing the drift in the delay. We took a series of 24 baseband data acquisitions, each 10~min apart, half at night and half during sunrise of 2023 June 4, and beamformed each toward the source J0117+892 from the VLBA Radio Fundamental Catalog (hereafter the ``North Celestial Pole'' or ``NCP'' source). The high declination of this source ensures that it is always in the field of view, making it an ideal test target.

For each set of 12 acquisitions, the first is taken as a phase calibrator. We then fit the fringes of the phase-referenced visibilities using the log-likelihood function in Equation~\ref{eqn:loglikelihood}, obtaining a delay $\tau$ for each sample. We performed this fit in two cases -- one assuming $\Delta$TEC=0 and another fitting for a non-zero $\Delta$TEC. Lastly, we subtract the calculated clock correction $\tilde{\tau}_{\text{clock}}$ from $\tau$ to get a residual delay $\tau_{\text{corrected}}$.

The results are shown in Figure~\ref{f:clock_iono}. The top panel shows the $\tau_{\text{corrected}}$ for the $\Delta$TEC=0 and the $\Delta$TEC$\neq$0 cases. The bottom panel shows the Allan deviation obtained from these delays and $\tilde{\tau}_{\text{clock}}$, along with the Allan deviation specification of the GPS clock and the Rb clock. We observe that $\tilde{\tau}_{\text{clock}}$ is the dominant contributor to $\tau$, and removing it gives a $\tau_{\text{corrected}}\leq$4~ns. Moreover, the Allan deviation of the $\tau_{\text{corrected}}$ for the $\Delta$TEC$\neq$0 case is in excellent agreement with the Allan deviation of the Rb clock, implying that the residual delay arises from inaccuracies in the Rb clock. Nevertheless, the 4-ns error is within our error threshold of 5~ns, and so for this analysis we can safely use phase calibrators separated from the target in time by $\sim$10$^3$~s.

\subsection{Ionosphere Characteristics}
\label{s:ionosphere_char}
\begin{figure}
    \centering
    \includegraphics[width=1\textwidth]{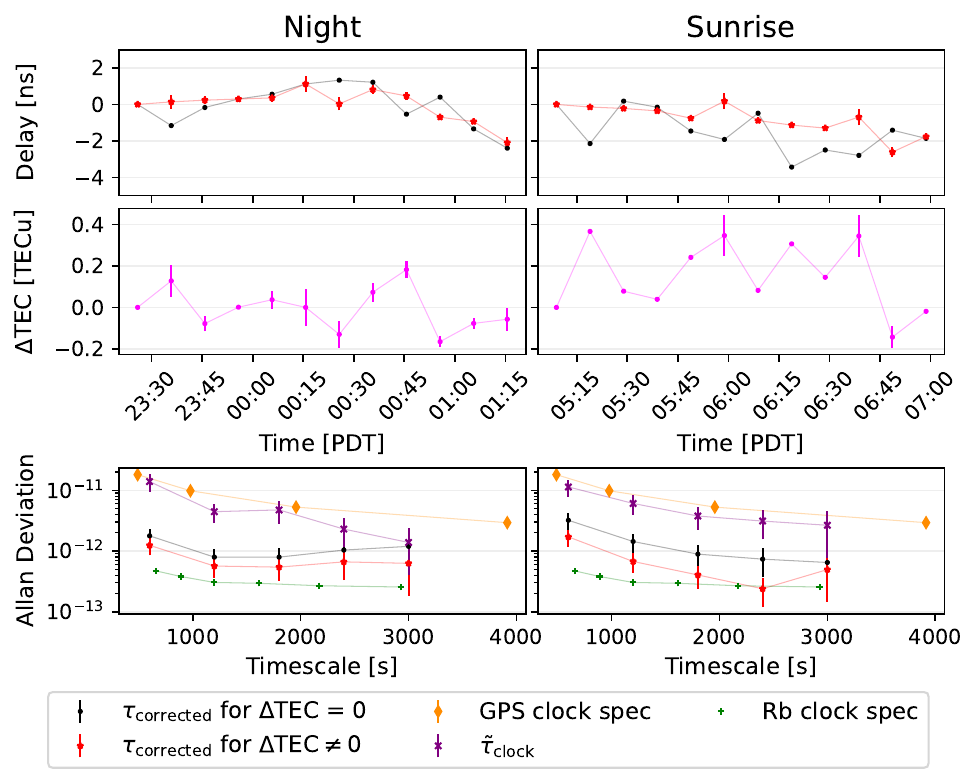}
    \caption{Plot showing the VLBI stability of the CHIME --- KKO baseline over a timescale of $\sim$10$^3$~s obtained using two NCP datasets having 12 acquisitions each and taken during night (left) and sunrise (right). For each of these datasets, the first acquisition is the phase calibrator for the rest of the acquisitions. The top panel shows the residual delay $\tau_{\text{corrected}}$ for the $\Delta$TEC=0 (black circles) and $\Delta$TEC$\neq$0 (red stars) cases. The $\tau_{\text{corrected}}$ are well within our error threshold of 5~ns. The middle panel shows the best-fit $\Delta$TEC, which, if not accounted for, can introduce an error of up to 2~ns in the residual delays. The bottom panel shows the Allan deviations obtained from the residual delays and $\tilde{\tau}_{\text{clock}}$ (purple crosses), along with the Allan deviation specification of the GPS clock (orange diamonds) and Rb clock (green pluses). Among the Allan deviations for the $\Delta$TEC=0 and $\Delta$TEC$\neq$0 cases, the $\Delta$TEC$\neq$0 case is more in agreement with the Allan deviation of the Rb clock. This suggests that fitting with $\Delta$TEC$\neq$0 results in a more accurate residual delay. 
    \label{f:clock_iono}}
\end{figure}

The two NCP datasets were used to estimate errors introduced in the VLBI delays due to the ionosphere when the TEC fluctuations are expected to be minimum (night) and maximum (sunrise).
The top panel of Figure~\ref{f:clock_iono} shows that for some acquisitions, there is a significant difference between the $\tau_{\text{corrected}}$ obtained for the $\Delta$TEC=0 and $\Delta$TEC$\neq$0 cases, implying a non-zero ionospheric contribution. The corresponding $\Delta$TEC values are shown in the middle panel of the figure. As expected, the magnitude of $\Delta$TEC and fluctuations in the ionosphere are more during sunrise.

We compared these best-fit $\Delta$TEC values with predictions of the International Reference Ionosphere (IRI, \cite{ISO16457}), as well as values interpolated from Global Ionosphere Maps (GIM) published by NASA/JPL\footnote{Obtained from the online archives of The Crustal Dynamics Data Information System (CDDIS), NASA Goddard Space Flight Center, Greenbelt, MD, USA}. Although we found good agreement between the IRI predictions and the GIM values, both were a factor of ten smaller than our measurements. We can account for this by considering that our VLBI measurements probe much smaller angular scales and higher TEC resolution than either the IRI or GIM. GIM maps are published with uncertainties on the scale of 0.1~TECu and at an angular resolution of 2.5$^\circ$ in latitude and 5$^\circ$ in longitude, while the CHIME and KKO sites are separated by $0.9^\circ$ in longitude. Fluctuations in the ionosphere, as caused by turbulence and medium-scale traveling ionosphere disturbances (TIDs), can be of order 10\% of TEC with wavelengths on the scale of 100~km \citep{coster_impact_2015}, and are especially strong during twilight hours.

The bottom panel of Figure \ref{f:clock_iono} shows that the Allan deviation for the $\Delta$TEC$\neq$0 case is in strong agreement with the Allan deviation of the Rb clock, proving that the best-fit $\Delta$TEC measurements are an accurate representation of the true ionospheric contribution. However, as shown by $\tau_{\text{corrected}}$ for the $\Delta$TEC=0 case, this contribution is not large enough to make our delay errors larger than 5~ns. We can thus safely choose to set $\Delta$TEC=0 for our FRB localizations on the CHIME --- KKO baseline.

\subsection{Baseline Calibration}
\label{sec:baseline}

To successfully correlate beamformed data between CHIME and KKO, we need to align the data streams in time to within the delay range of the correlator \citep{leung_andrew_pyfx}. This is achieved first by shifting the data arrays by a number of time frames (2.56~$\mu$s resolution), and then applying a phase to correct the remaining sub-frame difference. The applied phase correction depends on the baseline vector and the chosen pointing center, and so an error in the baseline vector will lead to a bias in the measured residual delay.

In this section, we will use visibilities formed on sources with known positions in order to compute the baseline vector. Although we have a rough approximation to the baseline vector from the GPS receivers at each site, the receiver locations are offset from the true phase centers of each station. The approximate baseline vector is sufficient for this analysis, which yields a correction to the baseline vector. We will demonstrate that this corrected baseline vector removes a clear bias in residual delays.

\subsubsection{Data Acquisition \& Target Selection}
We collected twenty-two 110-ms full-array baseband acquisitions at both CHIME and KKO over the course of three nights in late December of 2022, each triggered upon the detection of a giant pulse from the pulsar B0531+21 (the Crab pulsar) by the CHIME/FRB backend. Three continuum sources in the field, all largely separated from B0531+21, were chosen to be used for this baseline vector analyis: J054236.1+495107, PKS0531+19, and J0117+8928. These sources were originally identified during early commissioning after cross correlating a large sample of candidate sources in the VLBA Radio Fundamental Catalog, and were chosen because their cross-correlation signal is strong and their position are known to sub-arcsecond precision.

\subsubsection{Correlating \& Fringe-Fitting}

We assume fiducial positions for KKO and CHIME to perform the VLBI analysis. In the International Terrestrial Reference Frame (ITRF), these positions are:
\begin{align*}
	\tilde{\vecb{x}}_\text{KKO} &= (-2111738.254, -3581458.22, 4821611.987)~\mathrm{m} \\
	\tilde{\vecb{x}}_\text{CHIME} &= (-2059166.313, -3621302.972, 4814304.113)~\mathrm{m}
\end{align*}
The fiducial KKO position ($\tilde{\vecb{x}}_\text{KKO}$) comes from the GPS clock, which has a positional accuracy of a few meters and is located roughly 15~m from the center of the array. The CHIME position ($\tilde{\vecb{x}}_\text{CHIME}$) represents mean position of all active feeds in the CHIME array (the centroid), and has been used in other VLBI analyses with CHIME \citep{slb+23, lmm+21, cls+23, clr+21}.

The full-array baseband data from CHIME and KKO were beamformed toward the three continuum sources, and then correlated with \texttt{PyFX} \citep{leung_andrew_pyfx}. If either polarization failed to correlate with an S/N greater than 50, the visibilities for that source were not saved and the data excluded from the analysis. Out of the $22$ baseband dumps, $11$ were excluded because they did not meet the required S/N threshold. This was typical for events at the edges of the field of view where sensitivity is reduced. For all events, PKS0531+19 and J0117+8928 visibilities were divided by J054236.1+495107 visibilities. This in-beam \emph{phase-referencing} removes direction-independent instrumental and clock errors.

\begin{figure}
    \centering
    \includegraphics[width=1\textwidth]{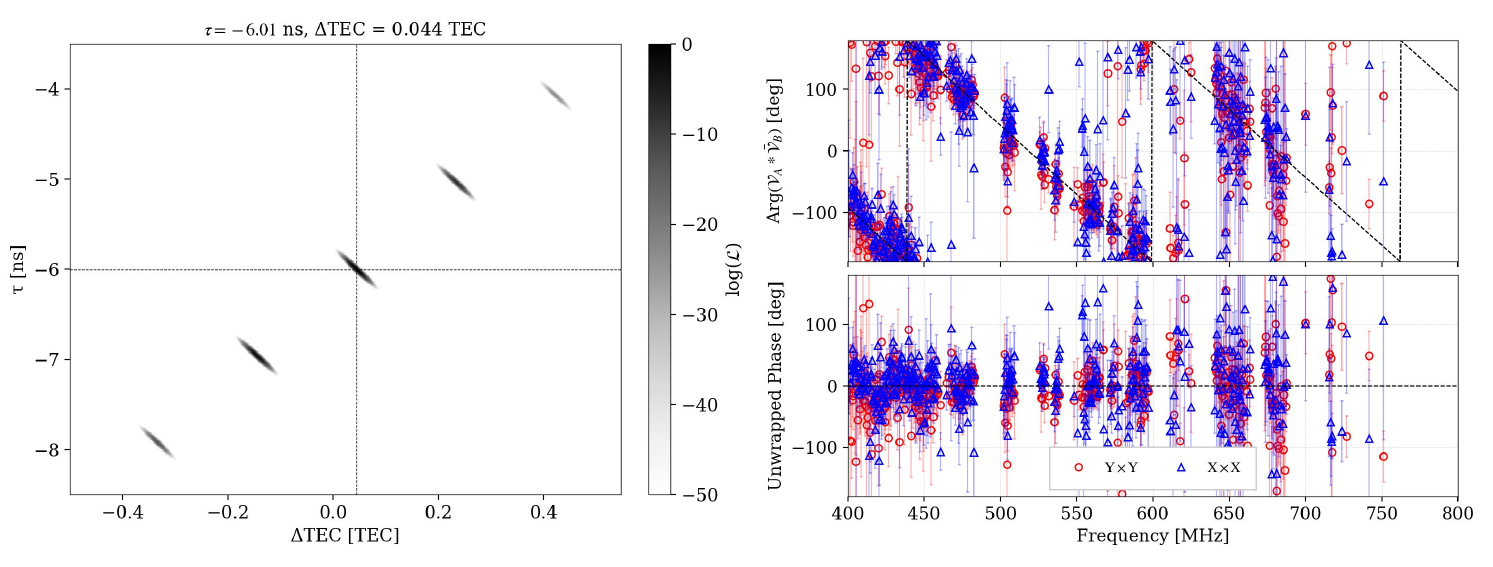}
    \caption{A sample fringe-fit of J0117+8928 phase calibrated by J054236.1+495107 to extract the non-dispersive residual delay arising from an incorrect baseline vector. \textit{Left:} Log-likelihood versus differential ionosphere density $\Delta$TEC and delay $\tau$, normalised to its maximum value and with color scale saturated to $\pm 5\sigma$. \textit{Right:} Cross-correlation visibility phases with $\mathrm{Y}\times\mathrm{Y}$ polarization in open red circles and $\mathrm{X}\times\mathrm{X}$ in open blue triangles. We apply $\tau$ and $\Delta$TEC corrections corresponding to the maximum likelihood to the phase and plot the original fit as well as the residuals.}
    \label{fig:fringe_fit_offset}
\end{figure}

We fit the phase model described in Equation~\ref{eq:phase_model} to the phase-referenced visibilities by maximizing the log-likelihood in Equation~\ref{eqn:loglikelihood}. For each target/calibrator pair, we perform an iterative grid search to identify the island of highest log-likelihood, marginalizing over $\Delta$TEC values. An example of one of the eleven fringe-fits performed on J0117+8928 is shown in Figure \ref{fig:fringe_fit_offset}. The left panel shows the log-likelihood grid search, normalized by subtracting the maximum value and saturating the color scale to $\pm 5\sigma$. The island of highest likelihood is indicated with the dashed-line crosshairs. The best-fit corresponding to this maximum value is shown on the right: $\mathrm{Y}\times\mathrm{Y}$ polarization provided by the open red circles and $\mathrm{X}\times\mathrm{X}$ polarization by the open blue triangles. The black dashed line indicates the best fit. No significant structure appears to be present in the residuals. 

\subsubsection{MCMC Fitting}\label{ss:offset_mcmc}
Fringe-fitting all phase-calibrated J0117+8928 and PKS0531+19 visibilities resulted in twenty-two delay measurements, each with an associated statistical error. If our assumed baseline vector station positions $\vecb{x}_\text{KKO}$ and $\vecb{x}_\text{CHIME}$ were perfectly correct, then the measured delays would all be zero, since the data were fringestopped to each source and calibrator position prior to phase-referencing. Let $\vecb{b} = \vecb{x}_\text{KKO} - \vecb{x}_\text{CHIME}$ be the true baseline vector and $\tilde{\vecb{b}} =  \tilde{\vecb{x}}_\text{KKO} - \tilde{\vecb{x}}_\text{CHIME}$ be the fiducial baseline vector used for fringestopping. The residual (non-dispersive) delay for a source, after fringestopping and phase referencing, is
\begin{equation}
	\label{eqn:residual_delay}
	\tau = \frac{(\vecb{b} - \tilde{\vecb{b}}) \cdot (\hat{n} - \hat{s})}{c},
\end{equation}
where $\hat{n}$ denotes the target source position and $\hat{s}$ the calibrator position. To estimate the baseline offset $\delta \vecb{b} = \vecb{b} - \tilde{\vecb{b}}$, we assume a log-likelihood function of the form
\begin{equation}\label{eq:log_llhd_mcmc}
	\text{log}\mathcal{L} = (\pmb{\tau} - \hat{\pmb{\tau}}(\delta \vecb{b}))^\text{T}\mathrm{C}^{-1}(\pmb{\tau} - \hat{\pmb{\tau}}(\delta \vecb{b})) - \text{log}(2\pi|\mathrm{C}|),
\end{equation}
and evaluate it with a Markov Chain Monte Carlo analysis over our sample of pulsar observations. Here $\pmb{\tau}$ is the vector of delays obtained from our fringe-fits, $\hat{\pmb{\tau}}$ represents the model delays (computed using \texttt{difxcalc11}) for a given baseline offset $\delta \vecb{b}$, and $\mathrm{C}$ is the covariance matrix. The full derivation of this covariance matrix is discussed in Appendix~\ref{app:delay_covar}. In short, it is designed to take into account both the statistical uncertainties in each measured $\tau$ as well as the systematic uncertainties due to the unknown primary beam phase. We note that if we ignore the systematic covariance effects in the noise matrix in Equation~\ref{eq:log_llhd_mcmc}, the mean delay after correlating the independent set of pulsars is biased to more negative delays and errors are underestimated. Introducing the noise model is required to remove this bias. Investigating the noise matrix can also provide insight into additional systematic contributions that are at play.

\begin{figure}
    \centering
    \includegraphics[width=1\textwidth]{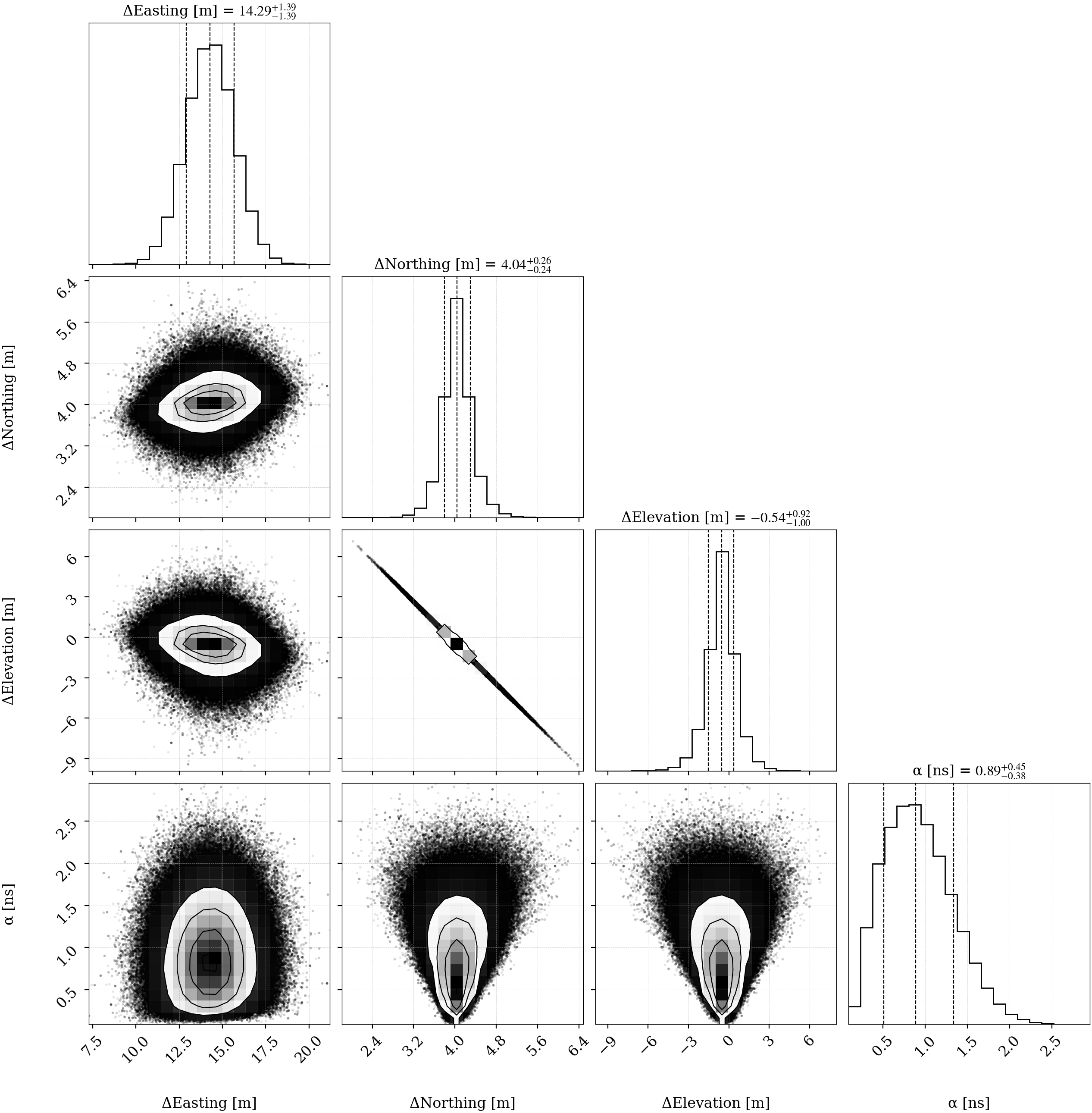}
    \caption{Baseline offset MCMC. Modeling twenty-two delay measurements using \texttt{difxcalc11}, we run a 150,000 step MCMC to determine the offset from the original position of the KKO GPS clock. The parameter $\alpha$ characterizes the amplitude of systematic errors in the delay measurements (see Appendix~\ref{app:delay_covar}). The log-likelihood is given by Equation~\ref{eq:log_llhd_mcmc}. No priors are assumed on any of the offset parameters. We assert a Gaussian prior on $\alpha$ with mean $1~\mathrm{ns}$ and width $0.4~\mathrm{ns}$, motivated by measurements of beam phase at CHIME (see Appendix~\ref{app:delay_covar} for details). The MCMC had a mean acceptance fraction of 55\% and the fit residuals are well behaved.}
    \label{fig:mcmc}
\end{figure}

The fit yielded a geocentric baseline offset of $\delta\vecb{b} = (-13.990^{+1.363}_{-1.364}, ~4.666^{+0.889}_{-0.908}, -1.906^{+0.595}_{-0.550})$~m at $1\sigma$ confidence interval. Converted to Easting, Northing and Elevation coordinates at the KKO site, the offset becomes $\delta\vecb{b} = (14.29^{+1.39}_{-1.39}, ~4.04^{+0.26}_{-0.24}, -0.54^{+0.92}_{-1.00})$~m, with $\alpha = 0.89^{+0.45}_{-0.38}$ nanoseconds (see Appendix~\ref{app:delay_covar}). Recall that this is a correction to a fiducial position taken at the GPS antenna on the receiver hut. This value is consistent with expectations, since the centroid of the array is located west and up from the receiver hut. Figure~\ref{fig:mcmc} shows a corner plot of the fitted baseline offset in Easting/Northing/Elevation coordinates. The MCMC had a mean acceptance rate of 55\% and the error bars on the quoted offsets include both statistical and systematic contributions.

If we assume that the fiducial CHIME position is exact, this baseline correction yields a KKO position of
\begin{align*}
	\vecb{x}_\text{KKO} &= \tilde{\vecb{x}}_\text{KKO} + \delta\vecb{b} \\
	 &= (-2111752.244^{+1.363}_{-1.364}, -3581453.556^{+0.889}_{-0.908}, 4821610.081^{+0.595}_{-0.550})~\mathrm{m}
\end{align*}
at MJD 59935. We note that positions will change over time due to tectonic motion, but the scale of this motion is negligible on this baseline at the required precision of our analysis.

\begin{figure}
    \centering
    \includegraphics[width=1\textwidth]{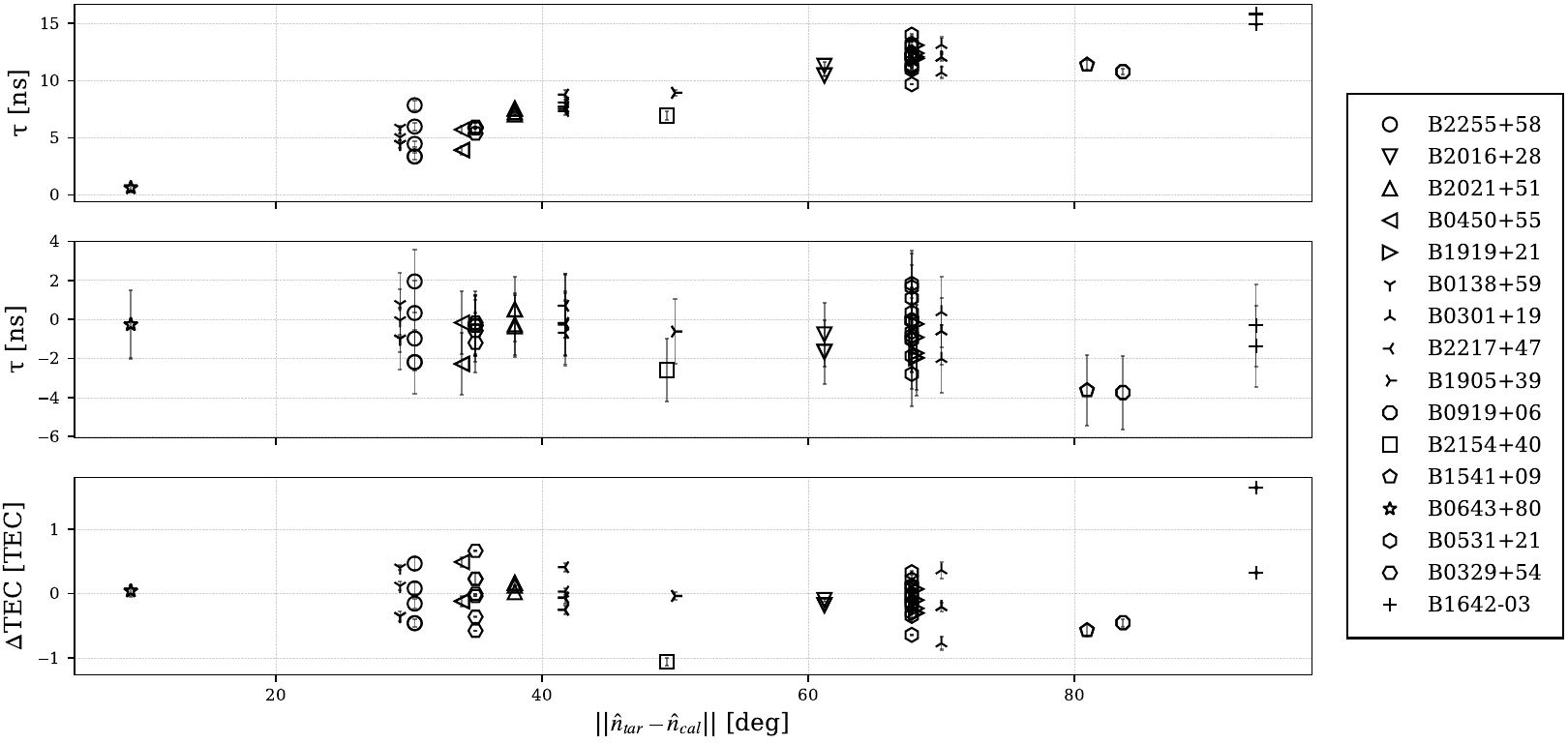}
    \caption{Extracted delays and TECs for a sample of 53 triggered baseband events from 16 unique pulsars referenced to J0117+8928 from June through August of 2023. \textit{Top Panel:} We plot the residual delay (nanoseconds) as a function of projected sky separation between target and calibrator (degrees) \textit{before} applying the baseline correction. Only statistical error bars are present which are clearly underestimated. \textit{Middle Panel:} For the identical set of pulsars, we plot the residual delays \textit{after} applying the offset correction. We compute the statistical and systematic uncertainties according to the covariance matrix derived in Appendix~\ref{app:delay_covar}. \textit{Lower Panel:} We plot the extracted TEC values from the fringe-fits along with their statistical error bars. We verify that the TEC values extracted before and after applying the correction were identical, confirming that our grid-spacing used to carry out the fringe-fit was sufficiently fine. The values have a mean and standard deviation of $0.0 \pm 0.4$ TECu.}
    \label{fig:pulsars_vs_ncp}
\end{figure}

To assess the validity of the fit, we test how the baseline correction affects measurements of a large sample of unique pulsars. We correlate, fringe-fit, and extract the residual delays of 53 pulsar events from 16 unique pulsars acquired from June to August of 2023, with and without applying the baseline correction. For each pulsar, we phase reference to the NCP source (J0117+8928) since it is always in the field of view of CHIME and KKO, allowing us to test a range of sky separations. The results are shown in Figure \ref{fig:pulsars_vs_ncp}. The top two panels show the residual delay (in nanoseconds) as a function of projected sky separation (in degrees) between the target and calibrator sources, where the top gives results without including the baseline offset, and the middle panel shows results including the baseline offset. There is a clear trend that residual delay increases with target--calibrator separation when the baseline vector is incorrect (top), as expected from Equation~\ref{eqn:residual_delay}. Including the computed baseline offset removes the trend (middle) to $\sim 1.5$~ns, even at extreme sky separations. The measured residual delays have a mean and standard deviation of $\bar{\tau} = -0.6 \pm 1.2~\mathrm{ns}$, suggesting that not all systematic contributions have been accounted for. However, the mean is biased by PSR B1541+09 and B0919+06, both of which have significant excess (negative) residual delay. We observe similar behaviour in the localization  of B0919+06 in Section~\ref{s:localization} and discuss the implications at the end of the section. In the bottom panel, we plot the extracted $\Delta$TEC values from the fringe fits with their statistical error bars. The fitted $\Delta$TEC values are unaffected by the baseline correction, indicating that fringe fitting grid had sufficiently fine spacing. Further, the fitted values, which largely all fall within $\Delta$TEC $\approx 0.0 \pm 0.4$~TECu, are small enough to be ignored when the target and calibrator are close together, and consistent with the results of Section~\ref{s:ionosphere_char}.

\subsection{Localization Performance}
\label{s:localization}

\begin{figure}   
\includegraphics[width=1\textwidth]{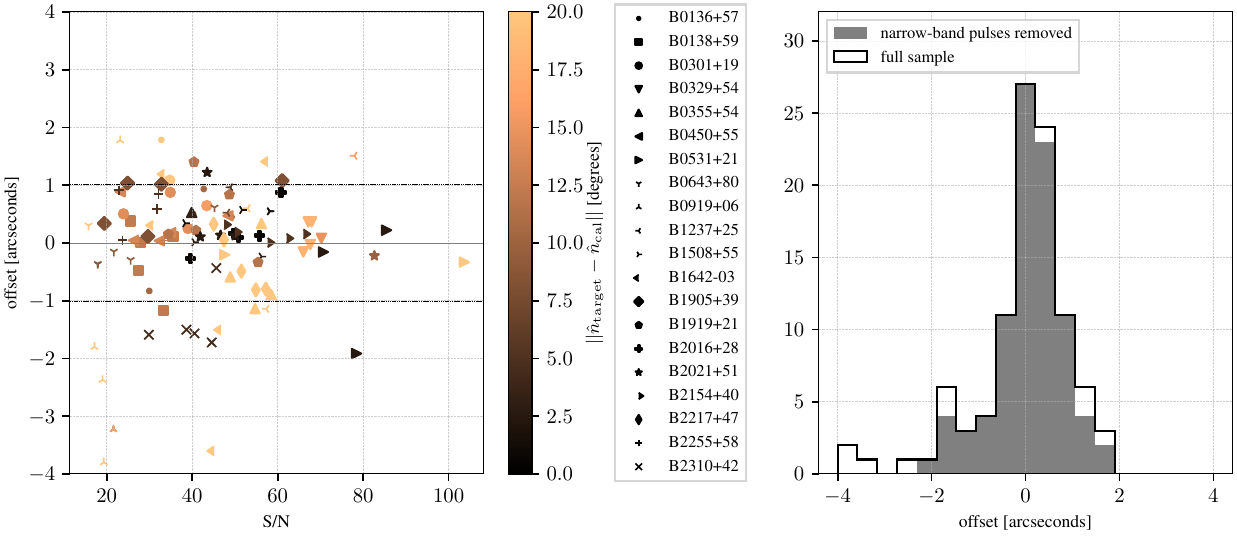}
    \caption{\label{f:loc_errors}Left panel: measured offsets of 100 pulses relative to positions obtained with the Very Long Baseline Array as a function of cross-correlation signal to noise. The points are color-coded by target-calibrator separation, and the root mean square of the entire sample is 1.16" (dashed horizontal lines). Right pane: Distribution of offsets for the entire sample (white outlined in black), and offsets after removing sources where the cross correlation signal could be measured in only half the band }
\end{figure}

The localization search begins with an initial guess for the position of an FRB given by the baseband localization pipeline, which uses CHIME's internal baselines and provides approximately arcminute level precision \citep{Michilli_2021}. After beamforming and correlating to this initial guess $\hat{n_{0}}$, and phase referencing to the closest in-beam calibrator from the VLBA Radio Fundamental Catalog, we evaluate the likelihood model Equation~\ref{eqn:loglikelihood} with the same same phase model $\phi(\nu,\Delta \hat{n})$ in Equation~\ref{eq:phase_model}, except with
\begin{equation}
    \Delta \hat{n} \equiv \hat{n}_{0}-\hat{n}
\end{equation}

where $\Delta \hat{n}$ denotes the difference between the initial pointing $\hat{n}_0$, to which the visibilities were fringe-stopped, and a given sky position $\hat{n}$. This provides a one-dimensional localization along the CHIME --- KKO baseline vector. To get a two-dimensional localization, we combine this with the $\sim 1'$ (CHIME-only) baseband localization.  The semi-major axis of our localization ellipse is perpendicular to the CHIME --- KKO baseline vector and is determined completely from the baseband localization pipeline (see also Figure 11 of \cite{leung_andrew_pyfx}). The semi-minor axis of the ellipse is obtained by taking the root mean square of a test sample of 20 unique pulsars, as shown in Figure~\ref{f:loc_errors}. Each pulsar was measured at 5 separate epochs, for a total of 100 localizations and a measured root mean square of 1.16$"$ on the offsets. Pulsar positions were then compared to the best determined position from a Very Long Baseline Array (VLBA) campaign (Proposal Codes: VLBA/21A-314, VLBA/22A-345, and VLBA/23A-099; \citealt{kaczmarek_vlba_prop_2020,curtin_vlba_prop_2022,curtin_vlba_prop_2023}). This campaign is measuring positions and proper motions for these pulsars for future calibration of the Outriggers using tracking beams \citep{pearlman_tracking_2024}. We note that the only three sources that lie beyond 3 $\sigma$ of our 1.16$"$ error budget are the two pulsars B1642-03 and B0919+06, which are far south in the CHIME field of view where the response is relatively weak. These three events had signal in only about one fourth of the band, which expectedly can lead to overfitting when trying to disentangle a geometric from an ionospheric phase contribution in Equation~\ref{eq:phase_model}. As a result, our 1.16$"$ error estimate will only be valid for FRBs with measured signal in over half the band.

\section{Conclusion and Outlook}
\label{s:conclusions}

Since KKO saw first light in June of 2022, it has passed through multiple stages of development to become a successful VLBI station, demonstrating the potential of the triggered-VLBI design of CHIME/FRB Outriggers. Through the commissioning process, we have thoroughly studied the $N^2$ visibility data in order to identify and correct defective components, measure feed positions, suppress and avoid RFI contamination, and compare system noise characteristics to CHIME. The set of pulsar and steady source observations, collected as triggered baseband dumps, has allowed us to test the latency of the triggering system, check the accuracy of the clock, fit for the ionosphere simultaneously with geometric delays, and better measure the baseline vector. Through the commissioning process we have also refined our monitoring procedures and ensured the system is robust against weather extremes and occasional power and Internet failures. With the conclusion of its commissioning phase, KKO is now under strict version control for its deployed systems and operating procedures. Nevertheless, we will continue to make improvements to ensure better reliability. We currently produce daily monitoring reports to provide a snapshot of data quality. For further monitoring, we plan to continue localizing known pulsars and start localizing repeating FRBs with known positions.

A key advantage to the KKO site is that, being so close to DRAO, the differential ionosphere density on the KKO --- CHIME baseline is generally negligible. From published ionosphere models we would expect the ionosphere contribution to the delay to be at a scale of 0.1~ns, which is well-within the 5~ns precision goal of this paper. However, we have demonstrated that we are capable of measuring the ionosphere contribution, and find values around 1.5~ns from fringe-fitting pulsar data for delay and $\Delta$TEC simultaneously (about 0.4 TECU at 600~MHz). Likewise, we find that including the ionosphere in fringe-fitting leaves us with an Allan deviation that matches the specifications for the rubidium clock, showing that our $\Delta$TEC fits are converging to sensible values. The observed ionosphere values are demonstrably negligible on our baseline, although noticeably larger than what published ionoshpere data would imply. Such published data capture variations in the ionosphere on larger angular and time scales, and are insufficient to predict the small-scale variations we are seeing. The ionosphere can have a highly variable contribution, and could conceivably impact our localization capabilities during periods of extreme solar activity, but our analysis here has demonstrated our ability to identify that impact via fringe-fitting. We will continue to monitor the ionospheric impact by localizing known pulsars.

From this initial work with pulsar data, we found that fringe-finding and localization were successful for events with S/N $\gtrsim 25$. Between April and October of 2023, KKO collected 93 FRB events with simultaneous recordings at CHIME, 43 of which had an S/N above 25. From April to the end of August, the \texttt{bonsai} threshold to trigger a baseband writeout at outriggers was set to 30, but after August 31st it was lowered to 15. Since then, there have been on average 13.2 FRB-triggers to KKO per week, with 4.4 FRB triggers/week above S/N 25. These statistics suggest that we will identify around 200 new FRB host galaxies in the next year, if we assume that each VLBI localization results in an unambiguous host association. This is not a fair assumption with the asymmetric localization region achieved with CHIME --- KKO alone, but with additional outriggers the localization region will be tightened further and associations will be much more obvious. We anticipate the full CHIME/FRB Outriggers VLBI network to be complete and operating by the end of 2024.

\section{Acknowledgements}

Funding for the CHIME/FRB Outrigger program is provided by a grant from the Gordon \& Betty Moore Foundation. We acknowledge that CHIME is located on the traditional, ancestral, and unceded territory of the Syilx/Okanagan people. We are grateful to the staff of the Dominion Radio Astrophysical Observatory, which is operated by the National Research Council of Canada.  CHIME is funded by a grant from the Canada Foundation for Innovation (CFI) 2012 Leading Edge Fund (Project 31170) and by contributions from the provinces of British Columbia, Qu\'{e}bec and Ontario. The CHIME/FRB Project, which enabled development in common with the CHIME/Pulsar instrument, is funded by a grant from the CFI 2015 Innovation Fund (Project 33213) and by contributions from the provinces of British Columbia and Qu\'{e}bec, and by the Dunlap Institute for Astronomy and Astrophysics at the University of Toronto. Additional support was provided by the Canadian Institute for Advanced Research (CIFAR), McGill University and the McGill Space Institute thanks to the Trottier Family Foundation, and the University of British Columbia. The CHIME/Pulsar instrument hardware was funded by NSERC RTI-1 grant EQPEQ 458893-2014. This research was enabled in part by support provided by the BC Digital Research Infrastructure Group and the Digital Research Alliance of Canada (alliancecan.ca).

\allacks

\bibliography{refs,frbrefs,psrrefs}

\appendix
\section{Delay Covariance}
\label{app:delay_covar}

In this appendix, we derive a form for the covariance matrix used in Section~\ref{ss:offset_mcmc} to estimate the baseline offset from measured residual delays. Recall that we started with 22 baseband acquisitions, then excluded 11 of these when any of the continuum sources failed to show a strong correlation. For each of the remaining acquisitions, we have three visibilities -- one for each source. Two of these are treated as targets, such that their visibilities are each divided by the visibility of the third source, acting as a phase and delay calibrator. The calibrated visibilities are then fringe-fit to obtain delays.

Let $\delta \tau_i = \tau(\hat{n}_i) - \tau(\hat{s}_i)$ denote the $i^\text{th}$ residual delay (Equation~\ref{eqn:residual_delay}), where $i = 1, \ldots, 22$.\footnote{Note a change in notation -- in Section~\ref{sec:baseline} we let $\tau$ denote the residual delay between target and calibrator. Here, we denote residual delay with $\delta \tau$ so we can use $\tau$ to represent the geometric delay per-source.} $\tau(\hat{n}_i) = \tilde{\vecb{b}} \cdot \hat{n}_i / c$ is the geometric delay associated with the target pointing $\hat{n}_i$, and $\tau(\hat{s}_i)$ is similarly defined for calibrator pointing $\hat{s}_i$. Our dataset comprises the 22 $\delta \tau$ measurements, and the raw delays $\tau$ are not measured.

Residual delay measurements $\delta \tau_i$ and $\delta \tau_j$ for $i \neq j$ may be correlated in one of two ways:
\begin{enumerate}
	\item $\hat{s}_i = \hat{s}_j$ (separate targets in the same baseband acquisition, sharing a calibrator source)
	\item $\hat{n}_i$ and $\hat{n}_j$ are close in telescope coordinates, resulting in similar beam phases.
\end{enumerate}
The first occurs because the two targets in each baseband acquisition share a calibrator, and the second is a systematic effect related to the structure of the primary beam. We will consider each effect when deriving off-diagonal terms of the covariance.

The total covariance matrix may be decomposed into a statistical and a systematic component: $\mathrm{C} = \mathrm{C}^{\text{stat}} + \mathrm{C}^{\text{sys}}$. We start by calculating the statistical covariance matrix. Its general form is
\begin{align}
	\mathrm{C}^{\mathrm{stat}}_{ij} &= \mathrm{Cov}(\delta \tau_i, \delta \tau_j) \\
	&= \mathrm{Cov}(\tau(\hat{n}_i),\tau(\hat{n}_j)) - \mathrm{Cov}(\tau(\hat{n}_i),\tau(\hat{s}_j)) - \mathrm{Cov}(\tau(\hat{n}_j),\tau(\hat{s}_i)) + \mathrm{Cov}(\tau(\hat{s}_i),\tau(\hat{s}_j))\\
	&= 
	\mathrm{Cov}(\tau(\hat{n}_i),\tau(\hat{n}_j)) + \mathrm{Cov}(\tau(\hat{s}_i),\tau(\hat{s}_j)).
	 \label{eq:cov_stat}
\end{align}
The inner terms cancel before the last step because delay errors are uncorrelated between the targets and calibrators.

When $i = j$, the covariance is just the variance of the delay:
\begin{equation}
	\mathrm{Cov}(\tau(\hat{n}_i), \tau(\hat{n}_j)) = \begin{cases}
		\sigma^2_{\hat{n}_i}, &  i = j \\
		0, &  i \neq j
	\end{cases}
	\label{eq:cov_targ}
\end{equation}
The first term in Equation~\ref{eq:cov_stat} is always zero for $i\neq j$ (when considering the statistical covariance), but the second term can be nonzero when $\hat{s}_i = \hat{s}_j$. Hence,
\begin{equation}
	\mathrm{Cov}(\tau(\hat{s}_i),\tau(\hat{s}_j)) = \begin{cases}
		\sigma^2_{\hat{s}_i}, &  i = j \\
		\sigma_{\hat{s}_i} \sigma_{\hat{s}_j}, & \text{if observations $i$ and $j$ share a calibrator} \\
		0, &  \text{otherwise}
	\end{cases}
	\label{eq:cov_cal}
\end{equation}

We do not know the per-pointing uncertainties $\sigma_{\hat{n}_i}$, since the analysis is done using phase-referenced visibilities. However, we can estimate them from $\sigma^2_i$ -- the uncertainty in $\delta\tau_i$ obtained from fringe fitting. It is related to the per-pointing uncertainties via:
\begin{equation}
	\sigma^2_i = \sigma^2_{\hat{n}_i} + \sigma^2_{\hat{s}_i}.
	\label{eq:total_sigma}
\end{equation}
Consider that the uncertainty in a measured delay is inversely proportional to the square of the signal to noise ratio:
\begin{equation}
	\sigma_{\hat{n}_i} = \frac{1}{2\pi \text{BW}_\text{eff} (\SNR)_{\hat{n}_i}} \qquad 
    \sigma_{\hat{s}_i} = \frac{1}{2\pi \text{BW}_\text{eff} (\SNR)_{\hat{s}_i}}, \label{eq:rogers}
\end{equation}
where BW$_\text{eff}$ is an effective bandwidth that is position-independent (see \cite{mena-parra_clock_2022, rogers_very_1970}) and assumed to be constant between both the target and calibrator. This is a safe assumption provided that the target and calibrator cover roughly the same frequency band, which is true in our case.

Defining $\gamma = 1/\left(2\pi\text{BW}_{\text{eff}}\right)$ to simplify notation, we apply Equation~\ref{eq:rogers} to Equation~\ref{eq:total_sigma},
\begin{align}
    \sigma_i^2 &= \gamma^2 \left( \frac{1}{(\SNR)_{\hat{n}_i}^2} +\frac{1}{(\SNR)_{\hat{s}_i}^2} \right)\\
    \implies \gamma^2 &= \left(
	\frac{(\SNR)_{\hat{n}_i}^2 (\SNR)_{\hat{s}_i}^2}{(\SNR)_{\hat{n}_i}^2 +
		 (\SNR)_{\hat{s}_i}^2} 
	\right)\sigma_i^2.
\end{align}
Inserting this into  Equation~\ref{eq:rogers} for $\sigma_{\hat{s}_i}$, we then get
\begin{equation}
    \sigma_{\hat{s}_i}^2 = \frac{\gamma^2}{(\SNR)_{\hat{s}_i}^2} = \left(\frac{(\SNR)^2_{\hat{n}_i}}{(\SNR)_{\hat{n}_i}^2 + (\SNR)_{\hat{s}_i}^2}\right)\sigma_i^2,
\end{equation}
providing a direct way of computing $\sigma_{\hat{s}_i}$ from S/N ratios and the fringe-fit uncertainty.

Collecting these results together, diagonal terms of the statistical covariance matrix are 
\begin{equation}
	\mathrm{C}_{ii}^{\mathrm{stat}} = \sigma^2_{\hat{n}_i} + \sigma^2_{\hat{s}_i} = \sigma^2_i,
\end{equation}
as a result of Equations~\ref{eq:total_sigma}, \ref{eq:cov_targ}, and \ref{eq:cov_cal}. The off-diagonal terms are zero except when observations $i$ and $j$ share a calibrator within the \textit{same} baseband acquisition, in which case the term is
\begin{equation}
  \mathrm{C}_{ij}^{\mathrm{stat}} = \sigma_i \sigma_j \sqrt{
  \frac{(\SNR)^2_{\hat{n}_i}}{(\SNR)_{\hat{n}_i}^2 + (\SNR)_{\hat{s}_i}^2} 
  } \sqrt{
	\frac{(\SNR)^2_{\hat{n}_j}}{(\SNR)_{\hat{n}_j}^2 + (\SNR)_{\hat{s}_j}^2} 
 }
  \qquad \text{ for } i \neq j.
  \label{eq:statcov_offdiag}
\end{equation}

Thus, we obtain the full statistical covariance matrix using only measured quantities --- per-pointing S/N ratios and the uncertainties from fringe-fitting. Note that Equation~\ref{eq:statcov_offdiag} only applies when $\hat{s}_i = \hat{s}_j$ and $\hat{n}_i \neq \hat{n}_j$ (e.g. for the shared calibrator within the \textit{same} baseband dump). It \textit{does not} apply for $\hat{s}_i = \hat{s}_j$ between separate baseband acquisitions since the measurements are safely assumed to be (statistically) independent of one another.

Next, we consider the systematic covariance matrix, which may be written as 
\begin{align}
	\mathrm{C}^{\mathrm{syst}}_{ij} &= \mathrm{Cov}(\delta \tau_i, \delta \tau_j) \nonumber \\
	&= \mathrm{Cov}(\tau(\hat{n}_i),\tau(\hat{n}_j)) - \mathrm{Cov}(\tau(\hat{n}_i),\tau(\hat{s}_j)) - \mathrm{Cov}(\tau(\hat{n}_j),\tau(\hat{s}_i)) + \mathrm{Cov}(\tau(\hat{s}_i),\tau(\hat{s}_j))
	\label{eq:cov_syst}.
\end{align}
In this case, the middle terms do not necessarily go to zero, since systematic effects can create nonzero correlation between different pointings (in cases when calibrator and target are close by). In the end, however, these will vanish.

The main systematic we consider here is the \emph{beam phase} -- The primary beam of each feed has a direction-dependent phase that is not known, and hence not included in our fringe fit. The beam phase at CHIME was measured in the CHIME Overview paper, shown in Figure~15 there to be stable and flat within the main lobe but rapidly varying outside the main lobe. Consequently, when target or calibrator pointings fall outside of the main lobe, an unknown phase offset may be introduced, which will change the measured delay. We include a direction-dependent model of systematic uncertainty in the form of an inverse Gaussian, which captures this rapid increase outside the main lobe:
\begin{equation}
	\sigma_{\hat{n}} \equiv \sqrt{\mathrm{Cov}(\tau(\hat{n}_i),\tau(\hat{n}_i))} = \alpha \times \mathrm{exp}\left[\frac{1}{2}\left(\left(\frac{X_n}{w_X}\right)^2 + \left(\frac{Y_n}{w_Y}\right)^2\right)\right].
\end{equation}
Here, $(X_n, Y_n)$ is the position of the source in projected-beam coordinates, $w_X$ and $w_Y$ are width parameters chosen to approximate the dimensions of the CHIME primary beam, and $\alpha$ is a parameter to be fitted. The CHIME primary beam has a measured East/West full-width at half-maximum (FWHM) of $\sim2.5^\circ$ at $600~\mathrm{MHz}$ and an estimated North/South FWHM of $\sim120^\circ$ at $600~\mathrm{MHz}$ (CHIME Overview). We therefore set $w_X = 1.06^\circ$ and $w_Y = 50.95^\circ$, corresponding with these FWHM values.

This model gives us an estimate of the uncertainty in beam phase for each pointing. The terms of the covariance matrix, if we assumed that these uncertainties are unrelated, would then just be various products of $\sigma$'s. However, we know that beam phase is smooth on certain angular scales. Figure~15 in CHIME Overview shows that within approximately $\delta\theta\leq 1.35^\circ$ of the meridian (in the East/West direction), the behavior of the beam phase is largely uniform. Physically, this corresponds to the scale prior to which reflections within the CHIME and KKO cylinders become the dominant beam effect. For two target pointings $\hat{n}_i$ and $\hat{n}_j$ we assume that two uncertainties are correlated if the distance between the two sources $\delta\theta \leq 2\cdot1.35^\circ = 2.7^\circ$.  Writing the corresponding source positions as $(X_i, Y_i)$ and $(X_j, Y_j)$, in the projected beam coordinates, the terms of the covariance matrix are multiplied by a function
\begin{equation}
	\xi(\hat{n}_i,\hat{n}_j) = 
	\begin{cases}
		1,     & \text{if } \sqrt{(X_i - X_j)^2 + (Y_i - Y_j)^2} \leq \delta\theta\\
		0, & \text{otherwise}.
	\end{cases}
\end{equation}

Combining the above results with Equation~\ref{eq:cov_syst}, we obtain our final model of the systematic covariance matrix: 
\begin{align*}
	\mathrm{C}^{\mathrm{syst}}_{ij} &= \alpha^2\left(\sigma_{\hat{n}_i}\sigma_{\hat{n}_j}\xi(\hat{n}_i, \hat{n}_j) - \sigma_{\hat{n}_i}\sigma_{\hat{s}_j}\xi(\hat{n}_i, \hat{s}_j) - \sigma_{\hat{n}_j}\sigma_{\hat{s}_i}\xi(\hat{n}_j, \hat{s}_i) + \sigma_{\hat{s}_i}\sigma_{\hat{s}_j}\xi(\hat{s}_i, \hat{s}_j)\right)\\
	&= \alpha^2\left(\sigma_{\hat{n}_i}\sigma_{\hat{n}_j}\xi(\hat{n}_i, \hat{n}_j)+ \sigma_{\hat{s}_i}\sigma_{\hat{s}_j}\xi(\hat{s}_i, \hat{s}_j)\right),
\end{align*}
The two inner terms disappear given the large $\gtrsim 20^\circ$ sky separation between the targets and the calibrator. Since we have reason to believe that the beam phase could induce errors on the order of a nanosecond, we impose a Gaussian prior on $\alpha$ centered on $1~\mathrm{ns}$ with a width of $0.4~\mathrm{ns}$. The width corresponds to the largest observed induced beam phase which was on the order of 1.74 rad at 700 MHz, corresponding to a delay of $\sim 0.4~\mathrm{ns}$.

\end{document}